\documentclass[useAMS,usegraphicx,usenatbib]{mn2e}

\title{Predicted properties of RR Lyrae stars in the SDSS photometric system}

\author[M. Marconi, M. Cignoni, M. Di Criscienzo, V. Ripepi, F. Castelli,
I. Musella, A. Ruoppo]{M. Marconi$^{1}$\thanks{E-mail: marcella@na.astro.it},
M. Cignoni$^{1}$, M. Di Criscienzo$^{1,2}$, V.Ripepi$^{1}$, F. Castelli$^{3}$,
I. Musella$^{1}$, \newauthor A. Ruoppo$^{1,4}$\\ $^{1}$INAF-Osservatorio
Astronomico di Capodimonte,,Via Moiarello 16, I-80131, Napoli, Italy\\
$^{2}$Universit\`a di Torvergata, Via della Ricerca Scientifica 1, 00133 Roma,
Italy\\ $^{3}$INAF-Osservatorio Astronomico di Trieste, via Tiepolo 11, 34131
Trieste, Italy,\\ $^{4}$Universit\`a Federico II, Complesso Monte S.Angelo,
80126 Napoli, Italy\\ }

\begin{document}

\date{Accepted ???. Received ???; in original form ???}

\pagerange{\pageref{firstpage}--\pageref{lastpage}} \pubyear{2002}

\maketitle

\label{firstpage}

\begin{abstract}
The luminosities and effective temperatures, as well as the whole bolometric
lightcurves  of nonlinear convective RR Lyrae models with 0.0001$\le$Z$\le$0.006
 are transformed into the
SDSS photometric system. The obtained  \emph{ugriz} lightcurves, mean magnitudes and
colors, pulsation amplitudes and color-color loops  are shown and analytical
relations connecting pulsational to intrinsic stellar parameters, similarly to the ones currently 
used in the Johnson-Cousins filters,
are derived. Finally the behaviour in the color-color planes is compared with available
observations in the literature and possible systematic uncertainties affecting 
this comparison are  discussed.

\end{abstract}


\section{Introduction}
Among Population II radially pulsating stars, RR Lyrae play a relevant role
both as standard candles and as stellar property tracers.  They are bright
stars, easily identified thanks to their characteristic variability (light
curves, periods, and colors) and their luminosity span a narrow range.  They
usually show low ($Z\sim$0.0001) to intermediate ($Z\sim{0.006}$) metallicity, 
but they can reach solar abundances in the solar neighbourhood or in the
Galactic bulge.  The calibration of their absolute magnitude $M_V(RR)$ in
terms of the measured iron-to-hydrogen content [Fe/H] allows us to use these
variables to infer the distances of Galactic Globular Clusters (GGC) and
nearby galaxies, as well as to calibrate secondary distance indicators such as
the GC luminosity function (see \citealt{dicri06} and references
therein). In this context several observational and theoretical efforts have
been made in the last years to provide accurate evaluations of the
Mv(RR)-[Fe/H] relation, usually approximated as $M_V(RR)=\alpha+\beta[Fe/H]$
(see e.g. \citealt{capcast}, \citealt{cacc} and references therein), and to
derive relevant implications for the Pop. II distance scale and the age of
globular clusters.  The above characteristics make RR Lyrae superb probes of
the old stellar populations and studies of their pulsation properties have
provided much of our present knowledge of the structure, kinematics, and the
metal abundance distribution of the halo. Moreover, they have been adopted in
the recent literature as fundamental targets of surveys devoted to the
identification of specific populations and galactic substructures (see e.g. \citealt{wu}, \citealt{bro}, \citealt{viva04} and references therein).  In
particular, the Sloan Digital Survey (SDSS) for RR Lyrae (\citealt{ive00})
and the QUEST RR Lyrae survey (Vivas et al. 2001, 2004, 2006) detected the tidal
stream from the Sagittarius dSph galaxy and other density enhancements in the
halo that may be other tidal streams, supporting the idea that RR Lyrae
surveys are crucial to trace the merger history of the Milky Way.  In the
context of the VLT Survey Telescope (VST) GTO (see \citealt{alca}) we
have planned a survey (STREGA@VST, see \citealt{marc06}) devoted at the
exploration of the southern part of the Fornax stream (\citealt{lyn}, \citealt{dine}) and the tidal interaction of the involved satellite
galaxies and globular clusters with the Milky Way halo. To this purpose RR
Lyrae will be used as tracers of the oldest stellar populations through
multi-epoch observations in the SDSS \emph{g} and \emph{i} filters. The comparison with
theoretical period-luminosity-color, period-luminosity-amplitude and Wesenheit
relations (see e.g. \citealt{marc03}; \citealt{dicri04}) will provide information on the individual distances
and in turn on the spatial distribution of the investigated stellar system.
In this context, the addition of multi-epoch \emph{r} and single epoch \emph{u} exposures 
on selected fileds will enable us to derive further constraints on the intrinsic 
stellar parameters of RR-Lyrae stars.  However, in order to correctly compare model predictions with
observations all the predicted pulsation observables need to be transformed
into the SDSS photometric filters. To this purpose in this paper we use
updated model atmospheres to predict the pulsation properties of RR Lyrae
stars of different metal contents in the $\emph{u, g, r, i, z}$
bands\footnote{These are the filters that will be mounted on the VST.}.

The organization of the paper is the
following: in Section 2 we present the adopted pulsation models, in
Section 3 we discuss the procedures adopted to transform the
theoretical scenario into the SDSS filters and in Section 4 we
illustrate the pulsation observables in these bands. Finally in Section
5 we compare model predictions with SDSS RR Lyrae data available in the literature,
discussing possible systematic errors affecting the comparison. 
The Conclusions close the paper.

\section{The adopted pulsation models}

During the last few years we have been computing an extensive and detailed set
of nonlinear nonlocal time-dependent convective models for RR Lyrae stars,
spanning a wide range of physical parameters and chemical compositions (see
e.g. \citealt{boncap03}, \citealt{marc03}, \citealt{dicri04}).  
In this paper we concentrate on
models with metallicity between $Z=0.0001$ and $Z=0.006$, that is the typical
range for Galactic Globular Cluster RR Lyrae. 
\begin{table}
 \centering %\begin{minipage}{15cm}
 \caption[]{Model input parameters. \label{tab1}}
 \begin{tabular}{cccc}
\hline
 $Z$  &   $Y$  &   $M/M_{\odot}$ &  $\log L/ \log L_{\odot}$ \\
\hline
\hline
 0.0001   & 0.24    & 0.65 & 1.61 \\
 0.0001   & 0.24    & 0.70 & 1.72 \\
 0.0001   & 0.24    & 0.75 & 1.61 \\
 0.0001   & 0.24    & 0.75 & 1.72 \\
 0.0001   & 0.24    & 0.75 & 1.81 \\
 0.0001   & 0.24    & 0.80 & 1.72 \\
 0.0001   & 0.24    & 0.80 & 1.81 \\
 0.0001   & 0.24    & 0.80 & 1.91 \\
 0.0004   & 0.24    & 0.70 & 1.61 \\
 0.0004   & 0.24    & 0.70 & 1.72 \\
 0.0004   & 0.24    & 0.70 & 1.81 \\
 0.001    & 0.24    & 0.65 & 1.51 \\
 0.001    & 0.24    & 0.65 & 1.61 \\
 0.001    & 0.24    & 0.65 & 1.72 \\
 0.001    & 0.24    & 0.75 & 1.61 \\
 0.006    & 0.26   & 0.58 & 1.55 \\
 0.006    & 0.26   & 0.58 & 1.65 \\
 0.006    & 0.26   & 0.58 & 1.75 \\
 \hline
\end{tabular}
 %\end{minipage}
 \end{table}
The adopted metallicity, Helium
content and stellar parameters are reported in Table 1, whereas the pulsation
properties of these models in the Johnson-Cousins bands, as obtained by
adopting the static model atmospheres by \citet{castgrat},
as well as an extensive comparison with observed RR Lyrae, are discussed in
our previous papers (see e.g. \citealt{dicri04}, hereinafter D04, \citealt{boncap03}, \citealt{marc03}, \citealt{boncap}).
In particular D04 discuss in detail the dependence of pulsation 
properties on the adopted mixing lenght parameter ($\alpha=l/H_p$ where $l$ is the mixing lenght and $H_p$ the pressure height scale) 
that enter the turbolent-convective model to close the nonlinear system of dynamical and convective equations (see \citealt{bonostell}).
From this analysis the authors conclude that the standard value ($\alpha=1.5$) adopted in all their previous investigations,  well reproduces the behaviour of RR Lyrae in the blue
regions of the instability strip, whereas there are indications for an increasing $\alpha$ value (up to 2.0) as one moves toward the red edge.
In this paper we concentrate on the models computed with $\alpha=1.5$ and the stellar parameters reported in Table 1, but we will 
mention the effect, if any,  of a possible alpha increase.

\section{Transformation into the SDSS photometric system}
In order to obtain the pulsation observables of the investigated RR Lyrae
models in the SDSS bands we have transformed both the individual static
luminosities and effective temperatures and the predicted bolometric light
curves into the corresponding filters.  In particular, as the adoption of
linear transformations from one photometric system to another introduces
uncertainties (limited precision and strong dependence on the color range), we
directly build magnitudes and colors for our RR-Lyrae models by convolving
model atmosphere fluxes with SDSS transmission functions \footnote{The SDSS transmission curves are available at url http://www.sdss.org/dr3/instruments/imager/index.html.}. In general, the
calculation of the magnitudes in a given photometric system (see e.g. \citealt{gir2002}) involves the integral equation:
\begin{equation}
m_{S_\lambda} = -2.5\,\log\left(\frac{ \int_{\lambda_1}^{\lambda_2}\lambda
f_\lambda S_\lambda \mathrm{d}\lambda }{\int_{\lambda_1}^{\lambda_2}\lambda
f_{\lambda}^0 S_\lambda \mathrm{d}\lambda}\right)+m_0
\label{eq_photon}
\end{equation}
where $S_\lambda$ is the transmission function, $f_\lambda$ is the stellar
flux (that corresponds to model atmospheres of known ($T_{eff}, [M/H], \log g
$)), $f_{\lambda}^0$ the zero point reference flux and $m_0$ the zero point
reference magnitude. Then, the final absolute magnitudes are computed by the
knowledge of the stellar radius.
 
In the SDSS photometric system (an ABmag system), $m_0=0$ and
   the zero point reference spectrum is the absolute flux of Vega at 5480 A (\citealt{fuku}).

As for model atmospheres, in this paper we adopt the homogeneous set of updated
   ATLAS9 Kurucz model atmospheres and synthetic fluxes (new-ODF
   models)\footnote{Available at url http://kurucz.harvard.edu/grids.html or http://wwwuser.oat.ts.astro.it/castelli/grids.html } computed with a new
   set of Opacity Distribution Functions \citep{castelli03}. These
   calculations assume steady-state plane-parallel layer atmospheres, covering
   a metallicity range $[M/H]: 0.5, 0.2, 0.0, -0.5, -1.0, -1.5, -2.0\,\,
   \mathrm{and} -2.5$, both for $[\alpha/Fe]=0.0$ and $[\alpha/Fe]=0.4$,
   temperatures between 3500 to 50000 K and $\log g$ from 0.0 to 5.0.  The
   \citet{kur90} model atmospheres had some recognized problems, in particular
   for effective temperatures lower than 4500 K. One of the reasons was the
   lack of $\mathrm{H_2 O}$ in the line opacity calculations and the use of approximate
   line data for TiO and CN. The ATLAS9 grids of the new-ODF models were
   computed with updated solar abundances from \citet{greve98} and
   considering molecular line lists which include $\mathrm{H_2 O}$ molecular
   transitions and updated TiO and CN data. A comparison between broad-band
   synthetic colors for late type giants computed from the new-ODF models and
   from the PHOENIX/NextGen models has shown a remarkable agreement (Kucinskas
   et al. 2005, 2006) in spite of the PHOENIX/NextGen
   models are computed with more molecular species than in ATLAS9 and by
   assuming spherical simmetry.  This last hypothesis is very important when
   the extent of the atmosphere is comparable with the radius of the star.
 RR-Lyrae stars have generally temperatures higher than 4000 K,
   however, as a sanity check, for selected models we also {\bf computed}
   magnitudes and colors by convolving the light curves with the
   PHOENIX/NextGen synthetic spectra \footnote{Available at the web
   site ftp://ftp.hs.uni-hamburg.de/pub/outgoing/phoenix/GAIA/} (see
   Sect. 5).

\section{The pulsation observables in the SDSS photometric system}

The application of the procedure discussed in the previous Section allows us
to predict the pulsation observables and their behaviour as a function of the
model input parameters in the SDSS filters. In the following we explore in
detail the morphological features of light curves and the behaviour of mean magnitudes and colors,
the
topology of the instability strip and color-color loops, as well as the main relations between the
periods (and the amplitudes) of pulsation and the intrinsic stellar
parameters.
\subsection{The light curves}
\begin{figure*}
\includegraphics[width=11cm]{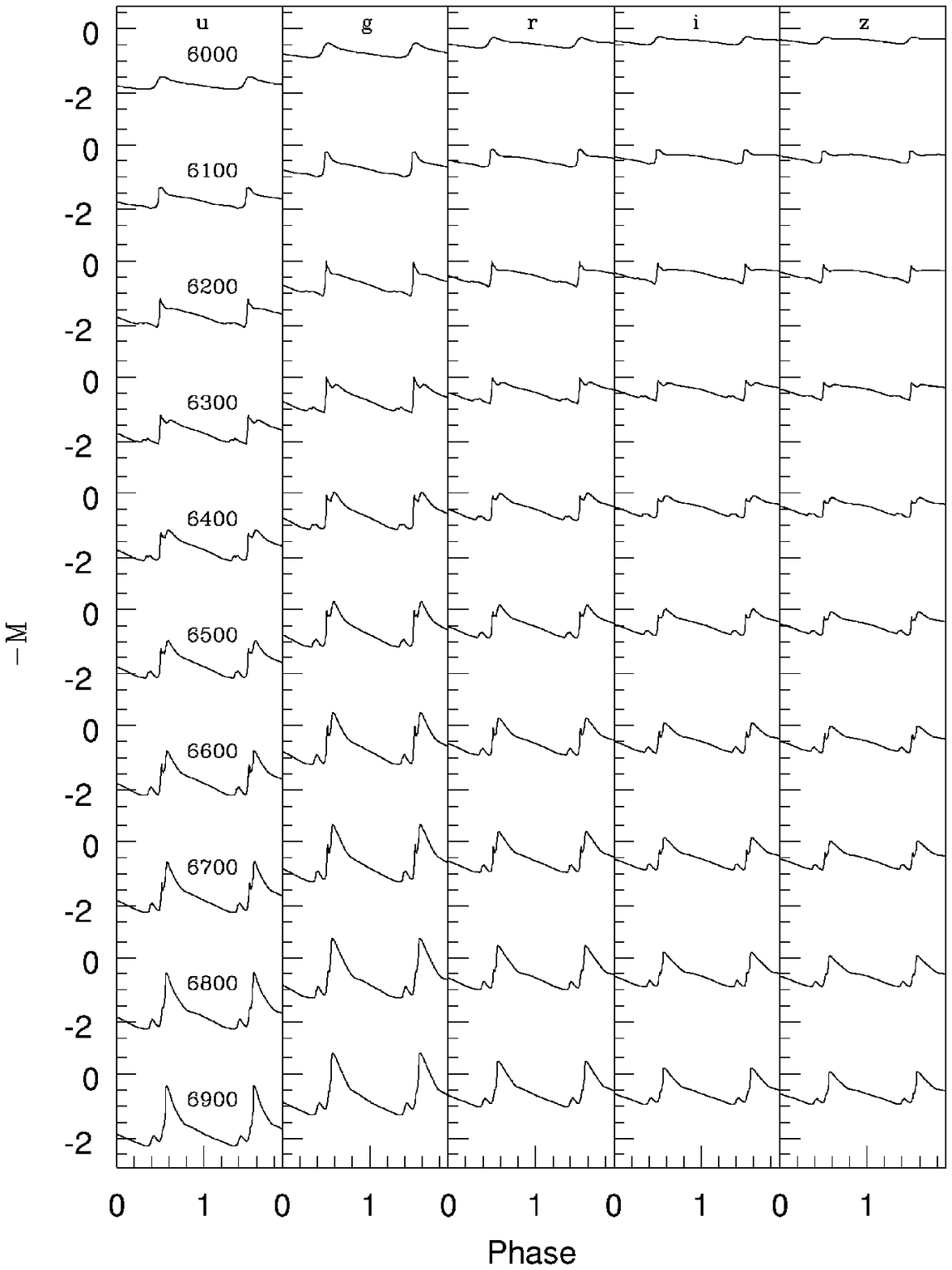}
\includegraphics[width=11cm]{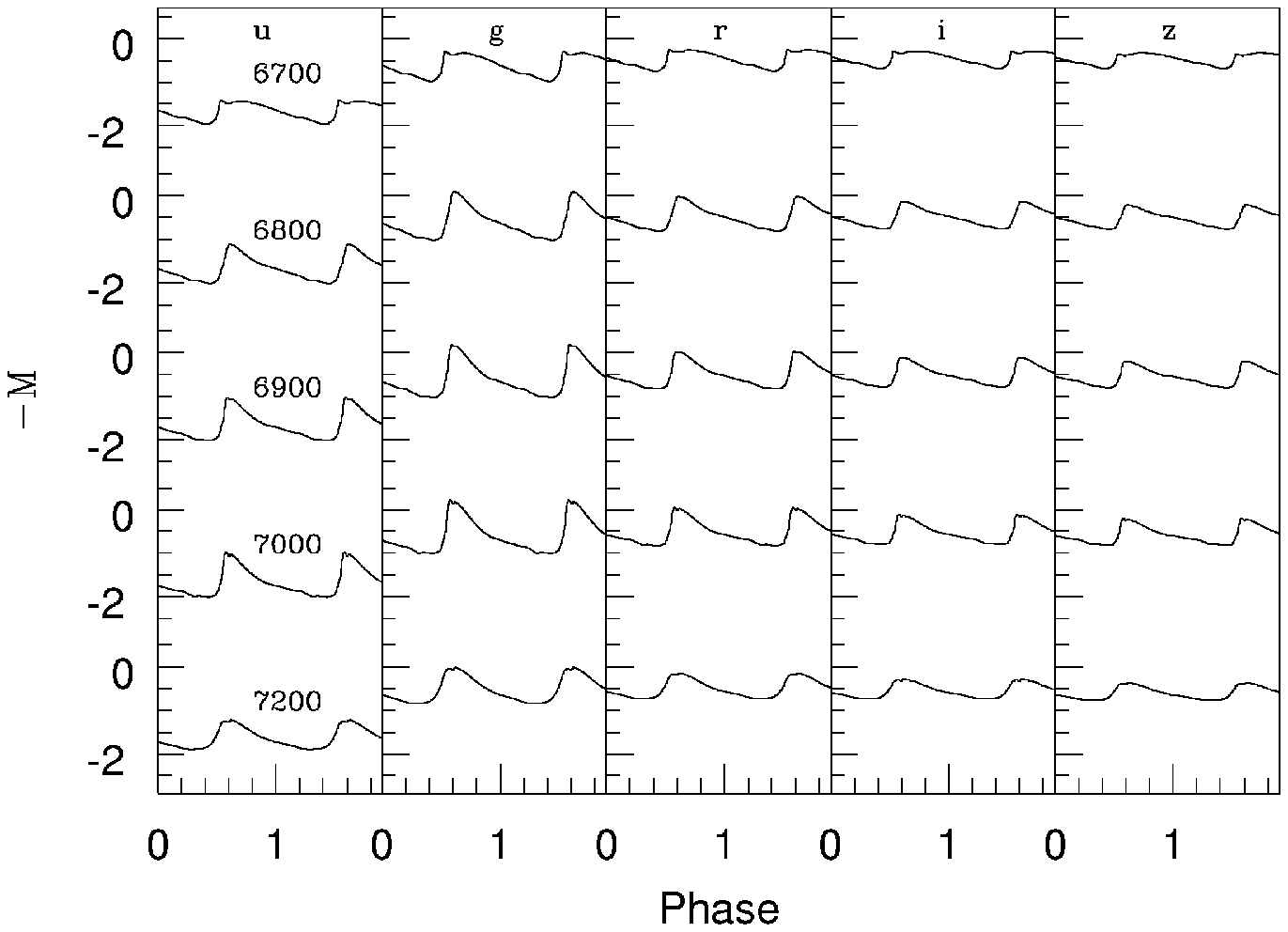}
\caption{Light curves trasformed in the labelled SDSS band for a number of
selected F(top) and FO-models(bottom) with Z=0.0001, $M=0.75M_{\odot}$,$ \log
L/ \log L_{\odot}=1.72$ at varying effective temperature}
\label{f1}
\end{figure*}

\begin{figure*}
\includegraphics[width=11cm]{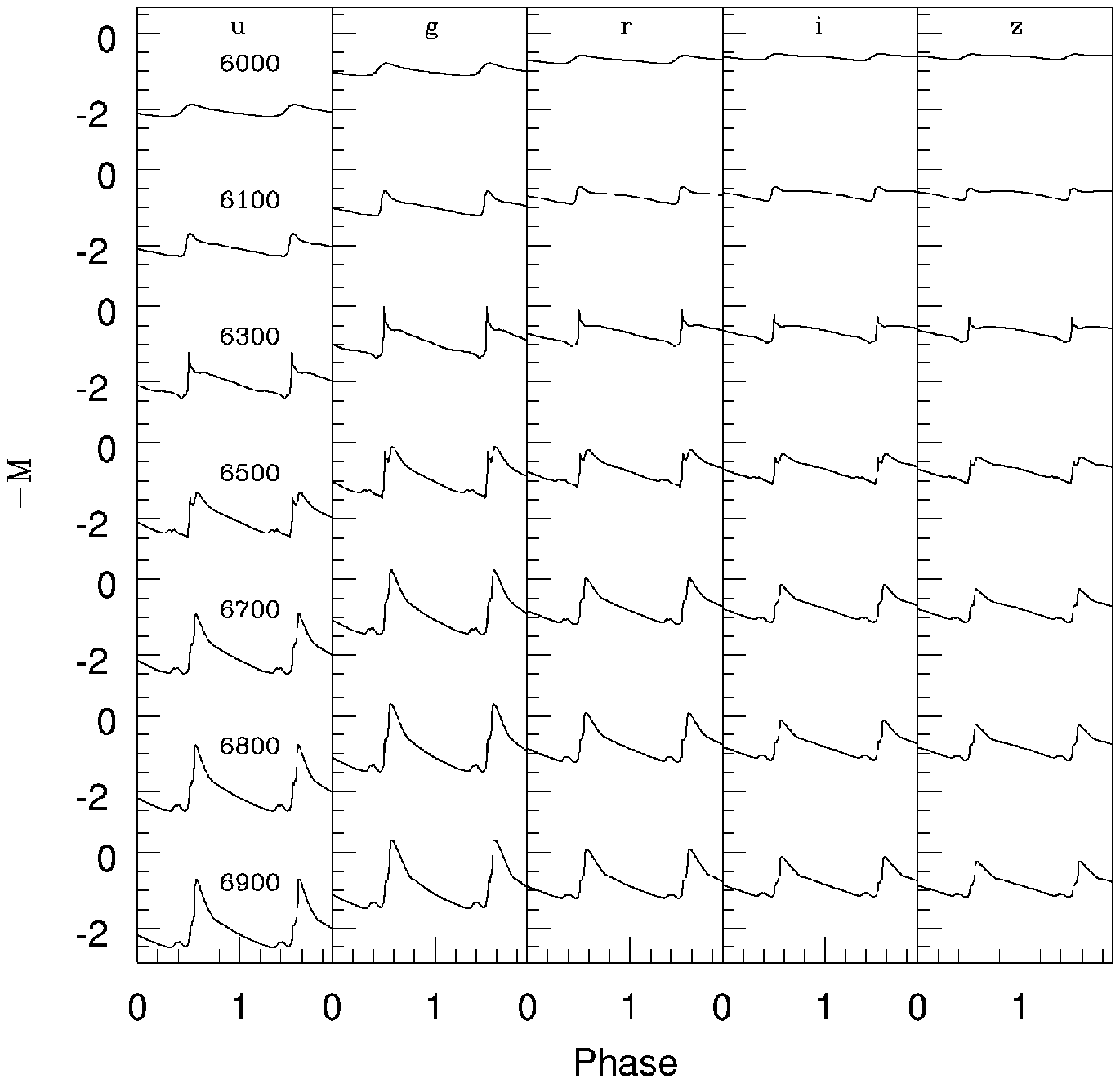}
\includegraphics[width=11cm]{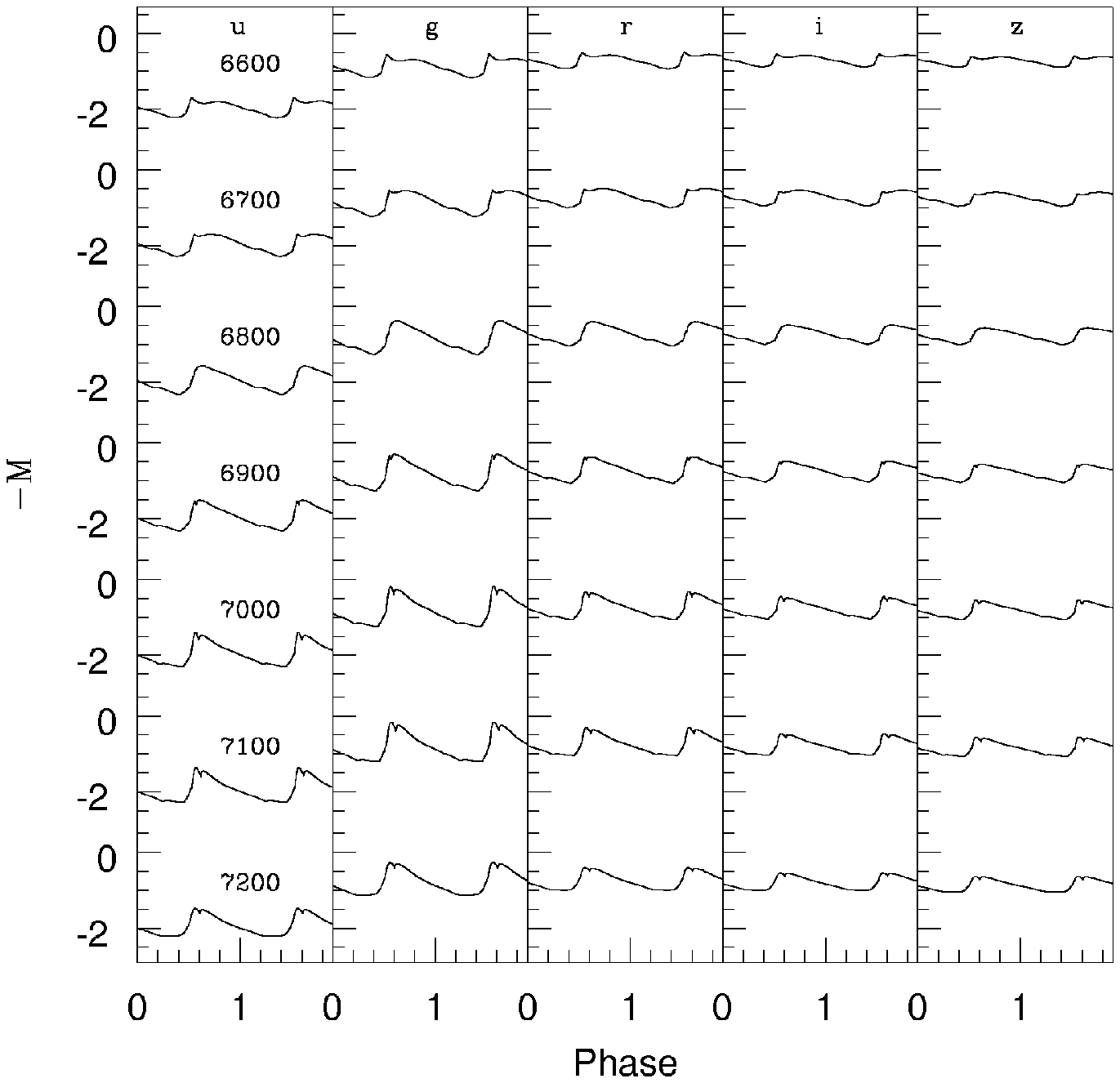}
\caption{Predicted SDSSS light curves for selected F (top) anf FO (bottom) models  with Z=0.001, $M=0.65M_{\odot}$,$ \log
L/ \log L_{\odot}=1.61$ and varying effective temperature.}
\label{f2}
\end{figure*}

\begin{figure*}
\includegraphics[width=12cm]{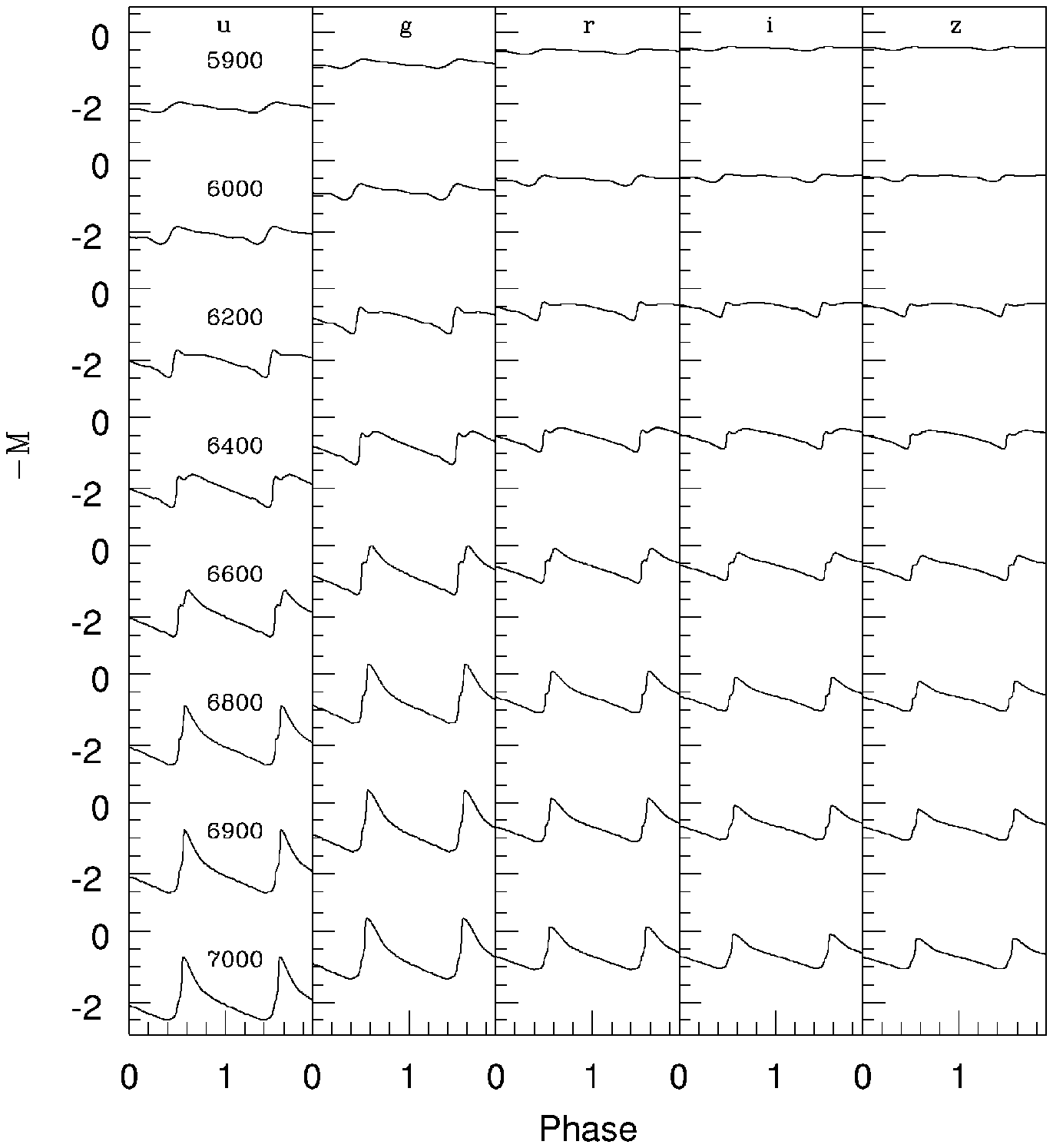}
\includegraphics[width=12cm]{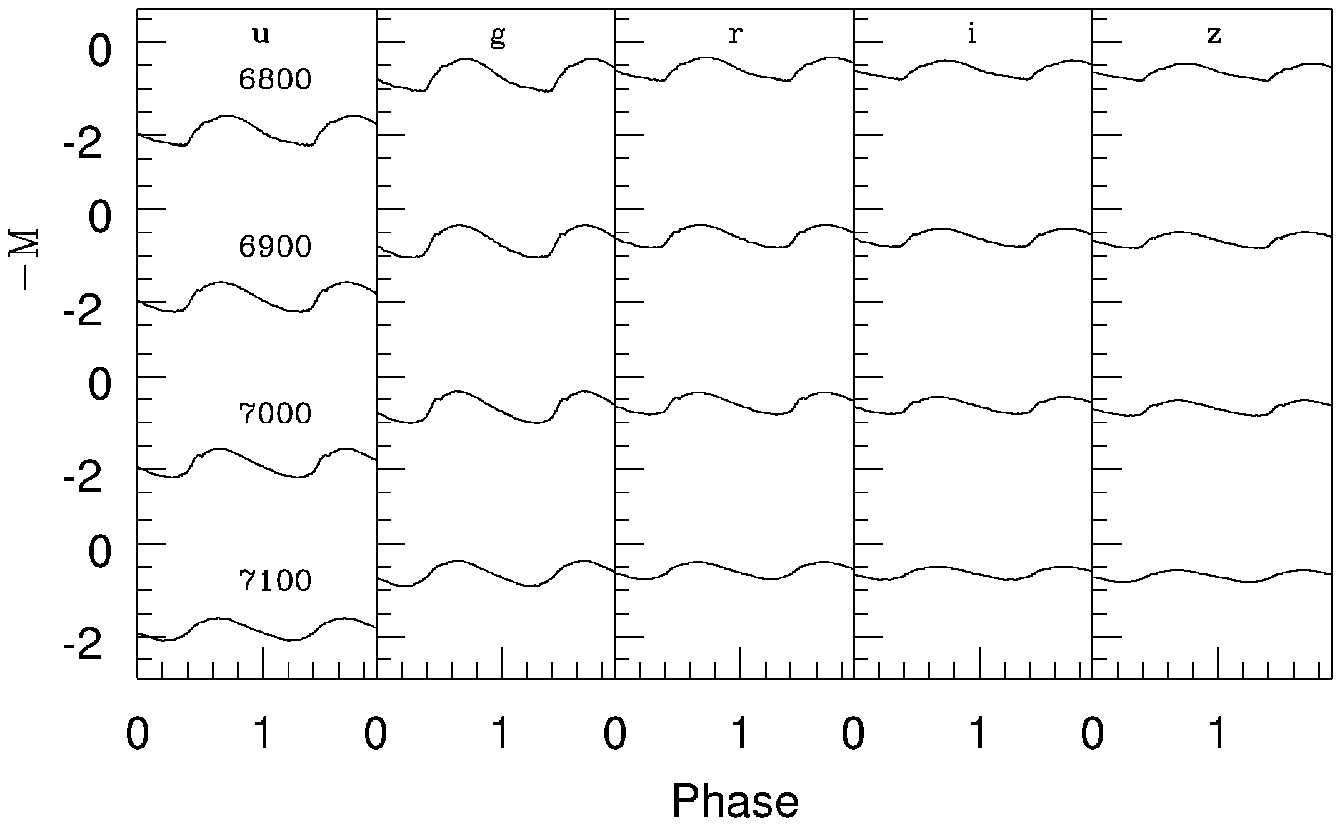}
\caption{Predicted SDSSS light curves for selected F (top) anf FO (bottom) models with Z=0.006, $M=0.58M_{\odot}$,$ \log
L/ \log L_{\odot}=1.65$ and varying effective temperature.}
\label{f3}
\end{figure*}

All the computed bolometric light curves have been transformed into the SDSS
 bands \footnote{The full set of transformed light curves is available upon
 request to the authors.}.  Figs. \ref{f1}-\ref{f3} show the transformed light
 curves for both fundamental {\bf(F)} and
 first overtone {\bf (FO)} pulsators at selected luminosity levels and $Z=0.0001, 0.001,
 0.006$.

 As already found for the Johnson-Cousins bands, the pulsation amplitudes vary with the wavelenght, 
increasing from  $u$ to  $g$  and decreasing from $g$ to $z$. 
Moreover, for a fixed filter,  the amplitudes of fundamental pulsators decrease from the blue
to the red edge, whereas  the first overtone ones  increase moving from
the blue edge to the middle of the instability strip and decrease as the red
edge is approached.
 The agreement between these predicted trends and the observed behaviour of both fundamental and first overtone pulsators has been extensively discussed 
in previous papers (see e.g. the discussion in \citet{bcr96} and  \citet{boncap}). We also notice that the theoretical 
light curves in the Johnson-Cousins bands
have been successfully compared with available  data in the literature (see \citet{bcm00}, \citet{mc05}). Unfortunately a similar comparison cannot be performed for the SDSS filters because of the lack of 
well-sampled RR Lyrae light curves in these bands.

\subsection{The mean magnitudes and colors}
\begin{figure}
\includegraphics[width=9cm]{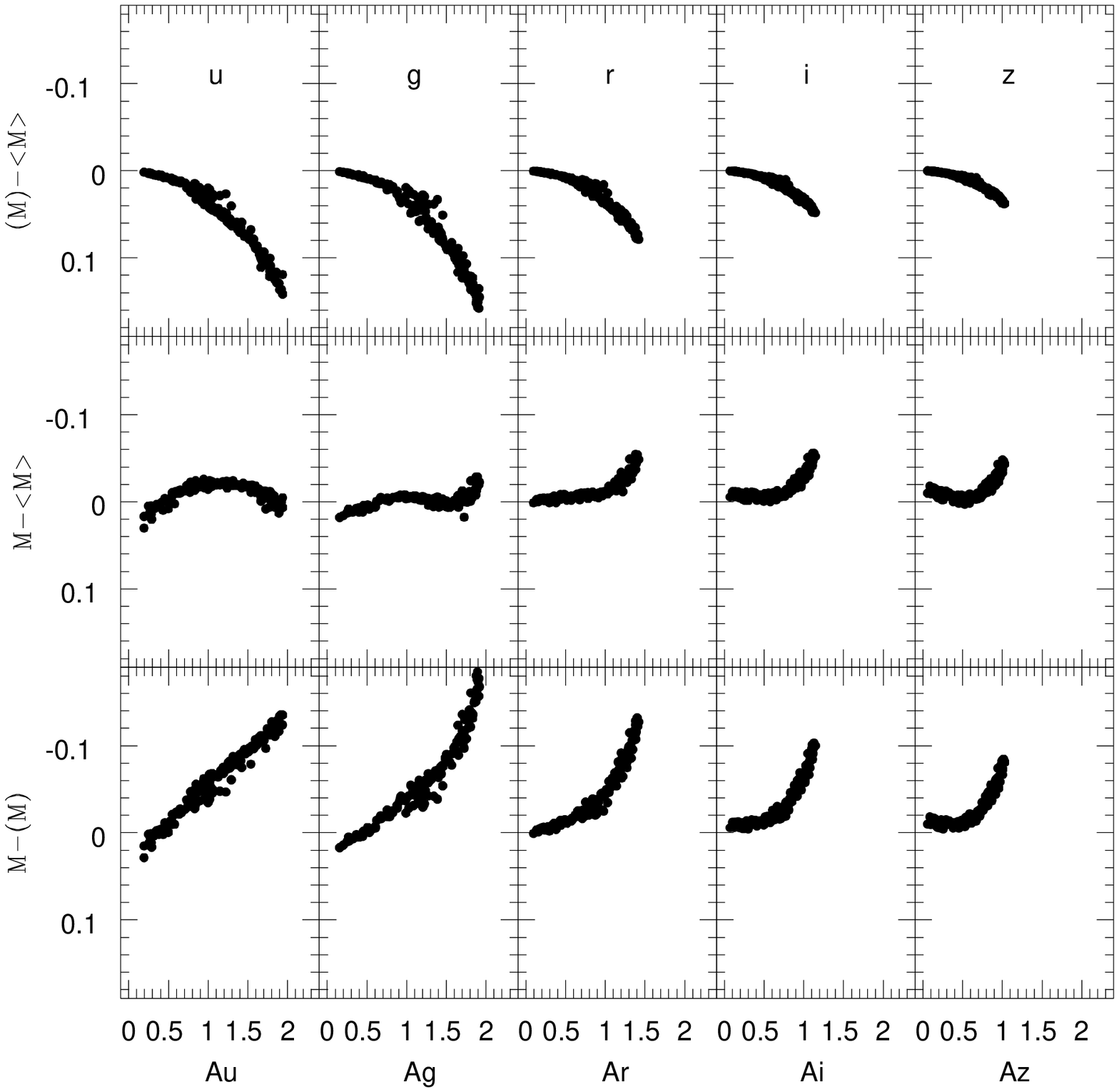}
\caption{Comparison between static and mean magnitudes for F-models.}
\label{f4}
\end{figure}

\begin{figure}
\includegraphics[width=9cm]{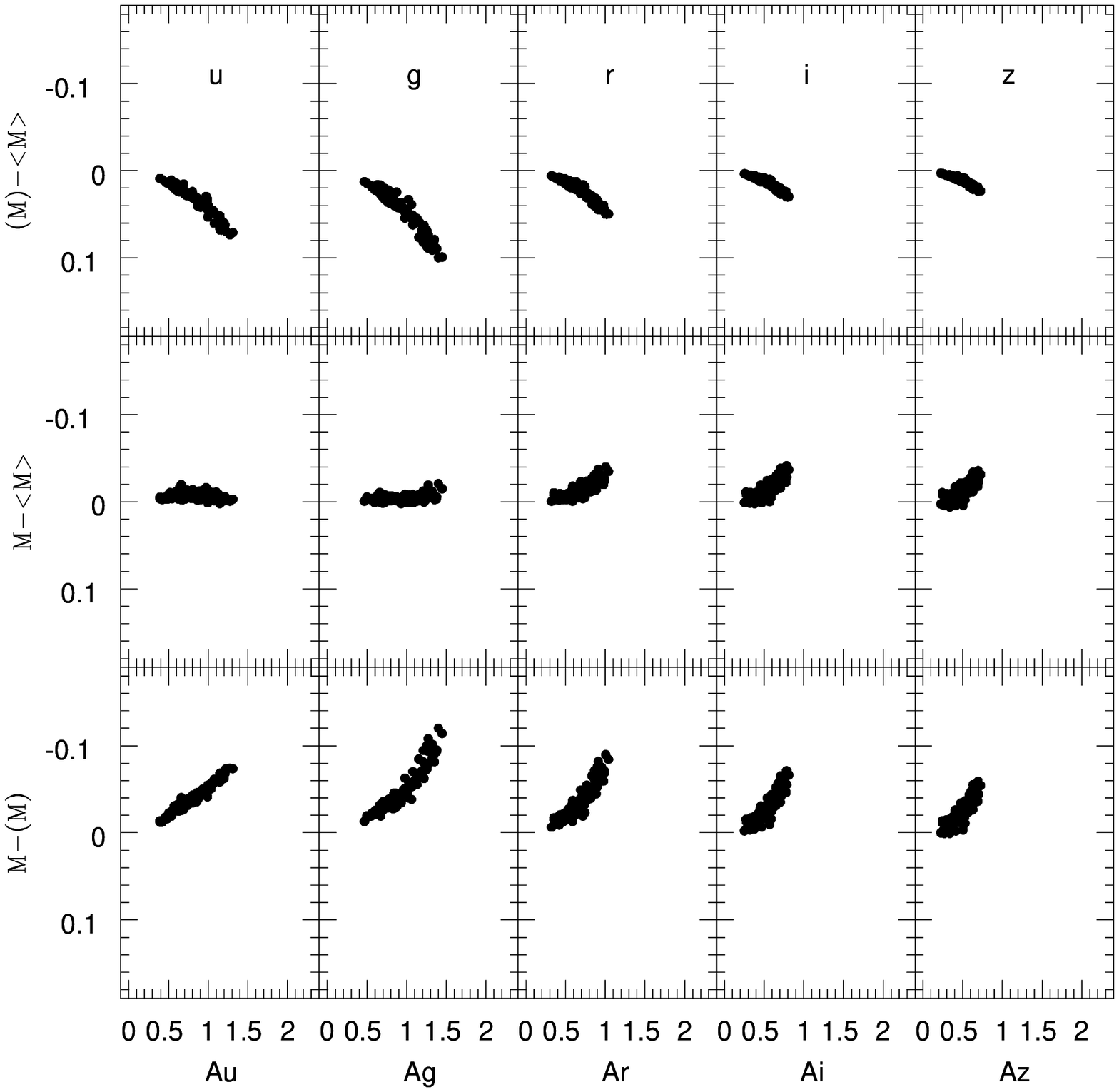}
\caption{The same as the previous figure but for FO-models.}
\label{f4a}
\end{figure}

From the obtained \emph{u, g, r, i, z} light curves we
can derive mean magnitudes and colors, following different averaging
procedures to produce either magnitude-averaged or intensity-averaged
values. As already extensively discussed for the Johnson-Cousins bands (see D04) the various types of mean values differ from the
static ones  the stars would have were they not pulsating. In Fig. \ref{f4}-\ref{f4a}
we show (for F and FO-models respectively) the difference between magnitude-averaged and intensity-averaged mean
magnitudes for the \emph{u, g, r, i, z} filters (top panels), as well as the
differences between the two kinds of mean values and the corresponding static
quantities (middle and bottom panels).

We notice
that similarly to what happens for the Johnson-Cousins filters, the difference
between magnitude-averaged and intensity-averaged magnitudes increases as the
corresponding pulsation amplitude increases, reaching the highest values in
the \emph{u} and \emph{g} bands. Moreover we confirm that intensity-averaged
magnitudes better reproduce the behaviour of static values than the
magnitude-averaged ones. Indeed the difference between the latter mean values and
the static ones can reach 0.2 mag in the \emph{g} band. For this reason in the following we will adopt 
the intensity-averaged mean
magnitudes in the \emph{u, g, r, i, z} filters. These quantities are  reported in Table 2 and 3 for
the whole fundamental and first overtone model sets respectively\footnote{Similar tables for
magnitude-averaged values are available upon request to the authors.}.
\begin{table*}
 \centering %\begin{minipage}{15cm}
 \caption[]{Intensity-averaged mean magnitudes for the full  set of fundamental  models (the full table is available in the electronic form).\label{tab2}}
 \begin{tabular}{cccccccccc}
\hline $Z$ & $M/M_{\odot}$ & $\log L/ \log L_{\odot}$ & $T_e$ (K) & P (d) &
 $<u>$ & $<g>$& $<r>$& $<i>$ & $<z>$ \\ \hline \hline 0.0001 & 0.65 & 1.61 &
 6900 & 0.4059 & 1.8978 & 0.8378 & 0.7813 & 0.8054 & 0.8419 \\ 0.0001 & 0.65 &
 1.61 & 6800 & 0.4260 & 1.9090 & 0.8445 & 0.7675 & 0.7808 & 0.8118 \\ ...  &
 ...  & ... & ... &... & ... & ....  & ... & ... & ... \\
  
\hline
 \end{tabular}                                                                            
 \end{table*}                                                                             
                                                                                         
\begin{table*}
\centering
\caption[]{Intensity-averaged mean magnitudes for the full set of first
overtone models (the full table is available in the electronic form). \label{tab3}}
\begin{tabular}{cccccccccc}
\hline $Z$ & $M/M_{ \odot}$ & $\log L/ \log L_{\odot}$ & $T_e$ (K) & P (d) & $
 <u>$ & $<g>$& $<r>$& $<i>$ & $<z>$ \\ \hline \hline 0.0001 & 0.65 & 1.61 &
 7300 & 0.2527 & 1.8838 & 0.7553 & 0.7572 & 0.8255 & 0.8918 \\ 0.0001 & 0.65 &
 1.61 & 7200 & 0.2634 & 1.8843 & 0.7737 & 0.7664 & 0.8242 & 0.8826 \\ ...  &
 ...  & ... & ... &... & ... & ....  & ... & ... & ... \\ \hline
\end{tabular}                                                                                           
\end{table*}      

\subsection{The instability strip boundaries}

\begin{figure}
\centerline{\includegraphics[width=8cm]{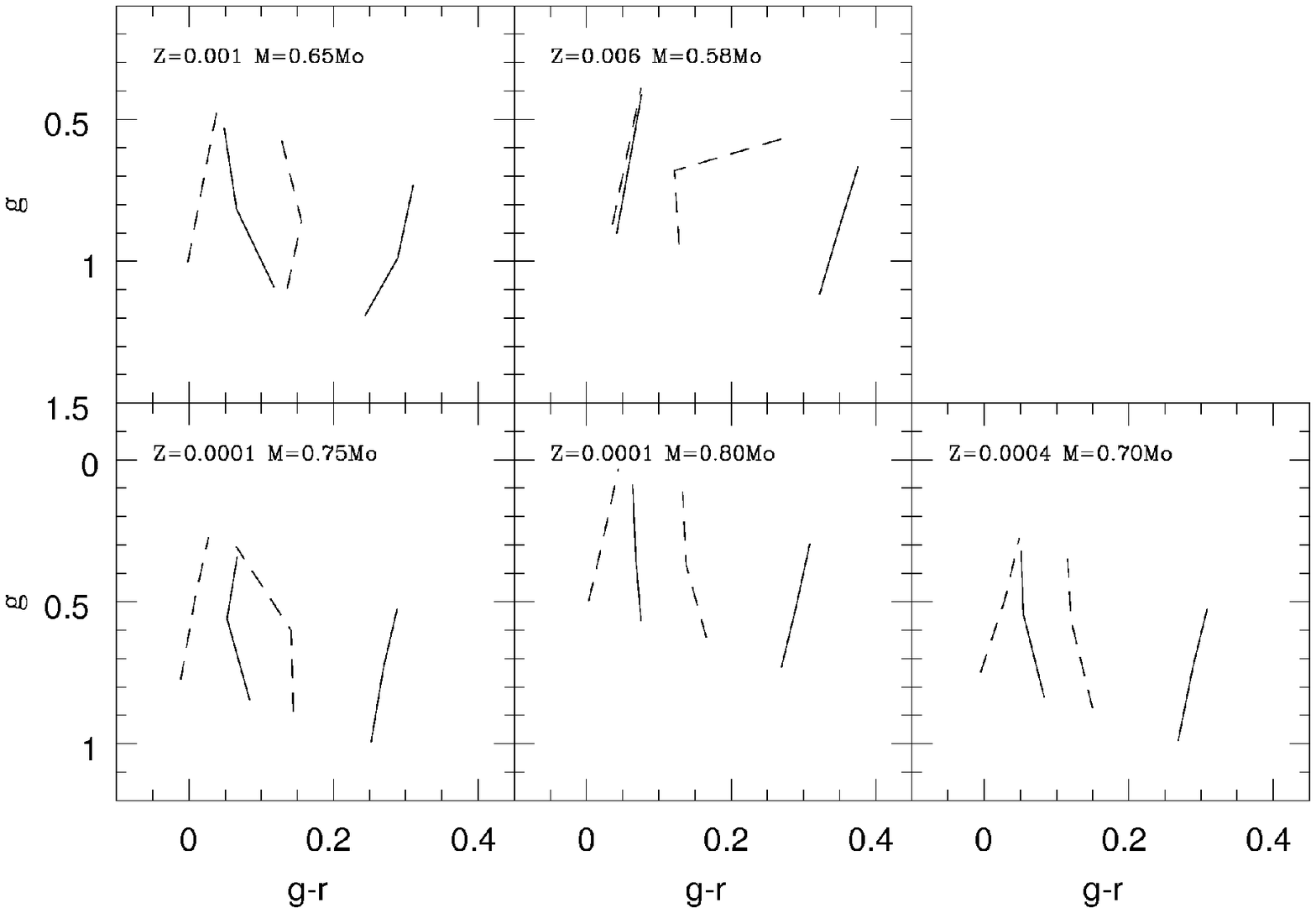}}
\caption{Predicted instability strip boundaries for both
F (solid lines) and FO (dashed lines) models in the intensity-averaged $\emph{g}$
versus $\emph{g-r}$ plane. The adopted metal abundances and stellar masses are labelld.}
\label{f5}
\end{figure}

In Fig. \ref{f5} we show the predicted instability strip boundaries for both
fundamental (solid lines) and first overtone (dashed lines) models in the intensity-averaged $\emph{g}$
versus $\emph{g-r}$ plane, for the metal abundances labelled in the different panels.
From left to right the different lines indicate the first overtone blue edge
(FOBE), the fundamental blue edge (FBE), the first overtone red edge (FORE)
and the fundamental red edge (FRE). At each magnitude level
blue (red) boundaries correspond to the effective temperatures of the first (last) pulsating model in the selected mode.  
Concerning the behaviour of the  boundaries  in
the magnitude-period diagram, a
linear regression to all the models quoted in Table 1 yields, by taking into
account the first (for the FOBE) and the last (for the FRE) pulsating models,
the mass-dependent analytical relations reported in Table 4.  
The zero point of the FOBE  relation vary by $\sim$0.08 mag in \emph{g} and by  $\sim$0.06 mag in \emph{i}
if the assumed $\alpha$ value increases from 1.5 to 2.0. The corresponding zero-point variation for the FRE
 is $\sim$0.3 mag in \emph{g} and  $\sim$0.01 mag in \emph{i} (see also the discussion in
 D04).
                                           
\begin{table}
 \centering
 %\begin{minipage}{15cm}
 \caption[]{Analytical relations for the FOBE and FRE in the
 form $M_i=a+b\log{P}+c\log{\frac{M}{M_{\odot}}}+d\log{Z}$. The dispersion $\sigma$ (mag) is also reported. 

\label{tab4}}
 \begin{tabular}{ccccccc}
\hline $M_i$ & $boundary$ & $a$ & $b$ & $c$ & $d$ & $\sigma$ \\ \hline \hline
 u & FOBE & 0.40 &-2.32 &-1.95 &0.086 &0.03\\ & FRE & 1.64 &-1.79 &-2.43 &0.12
 &0.03\\ g & FOBE &-1.04 &-2.43 &-2.14 & &0.04\\ & FRE & 0.18 &-1.99 &-2.34 &
 &0.02\\ r & FOBE & -1.12 & 2.63 &-1.85 & &0.03\\ & FRE & -0.07 &-2.23 &-1.92
 & &0.02\\ i & FOBE & -1.11 &-2.72 &-1.86 & &0.03\\ & FRE & -0.18 &-2.34
 &-1.96 & &0.01\\ z & FOBE & -1.09 &-2.78 &-1.91 & &0.02\\ & FRE & -0.22
 &-2.40 &-2.05 & &0.01\\ \hline\end{tabular} %\end{minipage}
 \end{table}

Only for the $u$
filter we also find a non negligible dependence on the metal content and in
all cases the standard deviations are of the order of few hundredths of
magnitude.

\subsection{The color-color loops}                      
                                        
\begin{figure}
\centerline{
\includegraphics[width=8cm]{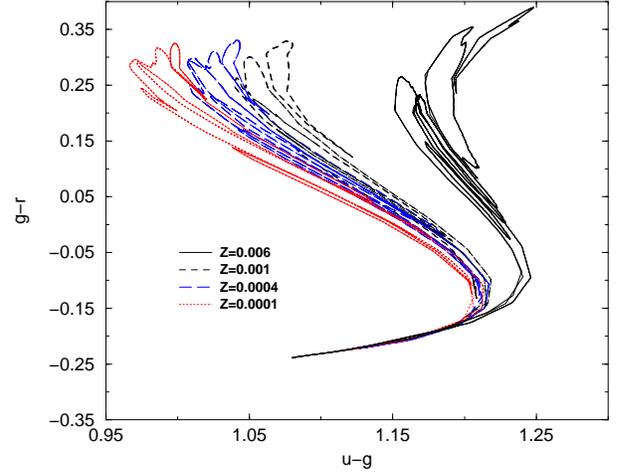}}
\caption{Theoretical loops in the \emph{g-r} vs \emph{u-g} diagram for selected
  models (see Table 5) at each adopted metallicity.}
\label{f6}
\end{figure}

\begin{figure}
\centerline{
\includegraphics[width=8cm]{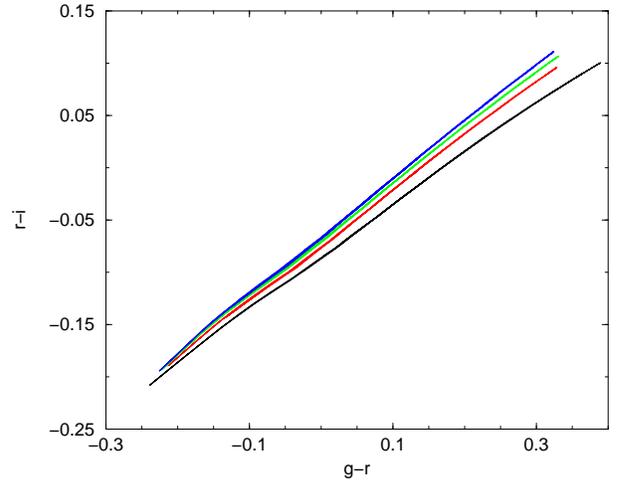}}
\caption{The same as in Fig. \ref{f6} but in the $r-i$ vs $g-r$ diagram; in this
  plane the loops move upward as the metallicity decreases. }
\label{f7}
\end{figure}

\begin{figure}
\centerline{
\includegraphics[width=8cm]{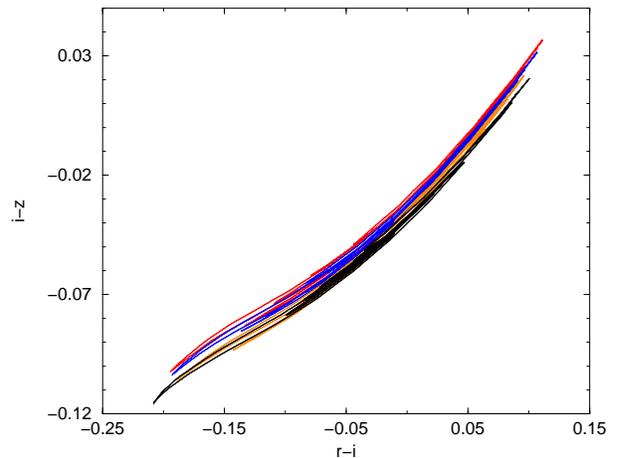}}
\caption{The same as in Fig. \ref{f6} but in the $i-z$ vs $r-i$ diagram}
\label{f8}
\end{figure}

The transformed multifilter light curves can be reported in different
color-color planes. In Figs. \ref{f6}-\ref{f8}  we show the theoretical loops in the \emph{g-r}
vs \emph{u-g},\emph{r-i} vs \emph{g-r} and \emph{i-z} vs \emph{r-i}  diagrams for selected
models (see Table 5) at each adopted metallicity. 
\begin{table}
 \centering
 %\begin{minipage}{15cm}
 \caption[]{Physical parameters of models plotted in the color-color diagrams. \label{tab5}}
 \begin{tabular}{cccccc}
\hline $Z$ & $Y$ & $M/M_{\odot}$ & $\log L/ \log L_{\odot}$ & $T_e$ & mode \\
 \hline \hline 0.0001 & 0.24 & 0.80 & 1.81 & 7100 & FO \\ 0.0001 & 0.24 & 0.80
 & 1.81 & 6800 & FO \\ 0.0001 & 0.24 & 0.80 & 1.81 & 6700 & F \\ 0.0001 & 0.24
 & 0.80 & 1.81 & 6300 & F \\ 0.0001 & 0.24 & 0.80 & 1.81 & 5900 & F \\ 0.0004
 & 0.24 & 0.70 & 1.72 & 7000 & FO \\ 0.0004 & 0.24 & 0.70 & 1.72 & 6700 & FO
 \\ 0.0004 & 0.24 & 0.70 & 1.72 & 6700 & F \\ 0.0004 & 0.24 & 0.70 & 1.72 &
 6400 & F \\ 0.0004 & 0.24 & 0.70 & 1.72 & 5950 & F \\ 0.001 & 0.24 & 0.65 &
 1.61 & 7200 & FO\\ 0.001 & 0.24 & 0.65 & 1.61 & 6800 & FO\\ 0.001 & 0.24 &
 0.65 & 1.61 & 6700 & F\\ 0.001 & 0.24 & 0.65 & 1.61 & 6400 & F\\ 0.001 & 0.24
 & 0.65 & 1.61 & 6100 & F\\ 0.006 & 0.26 & 0.58 & 1.65 &7100 & FO\\ 0.006 &
 0.26 & 0.58 & 1.65 &6900 & FO\\ 0.006 & 0.26 & 0.58 & 1.65 &6800 & FO\\ 0.006
 & 0.26 & 0.58 & 1.65 &7000 & F\\ 0.006 & 0.26 & 0.58 & 1.65 &6400 & F\\ 0.006
 & 0.26 & 0.58 & 1.65 &6000 & F\\ \hline
 \end{tabular}
 %\end{minipage}
 \end{table}

We notice that the metallicity effect is more evident in Fig. \ref{f6} due to the significant sensitivity of the \emph{u} band on metal abundance. On the other hand, in the $r-i$ vs $g-r$ plane the loops are very narrow and the effect of metallicity is smaller, so that, once the metallicity is known, the comparison between theory and observations in this plane could be used to evaluate color excesses. As for the  $i-z$ vs $r-i$ diagram, the predicted loops are very close to each other and the metallicity dependence is much less evident than for the other color combinations.
A linear regression through the intensity-averaged mean magnitudes reported in Tables 2 and 3 provides the following metal-dependent  analytical color-color relations:
\begin{eqnarray}
\lefteqn{u-g=1.41-0.12(g-r)+0.088\log{Z}\,\,\,\,\,(\sigma=0.03)}\\
\lefteqn{g-r=0.234+2.11(r-i)+0.035\log{Z}\,\,\,\,\,(\sigma=0.006)}\nonumber\\
\lefteqn{r-i=0.085+2.04(i-z)+0.0097\log{Z}\,\,\,\,(\sigma=0.004)}\nonumber
\end{eqnarray}
\noindent
holding for fundamental pulsators, and: 
\begin{eqnarray}
\lefteqn{u-g=1.35-0.19(g-r)+0.065\log{Z}\,\,\,\,\,\,\,(\sigma=0.02)}\\
\lefteqn{g-r=0.217+1.95(r-i)+0.0258\log{Z}\,\,\,\,\,(\sigma=0.006)}\nonumber\\
\lefteqn{r-i=0.082+2.00(i-z)+0.0062\log{Z\,\,\,\,\,(\sigma=0.003)}}\nonumber
\end{eqnarray}
\noindent
for the first overtone ones.

\subsection{The multifilter Period-Magnitude-Color relations}

The obvious outcome of the pulsation relation (see van Albada \& Baker 1971, Di Criscienzo et al. 2004) connecting the period to the mass, the luminosity and the effective temperature  into the observational plane is the Period-Magnitude-Color (PMC) relation, where the pulsation period for each given mass is correlated with the pulsator absolute magnitude and color. A linear regression through the intensity-averaged values reported in Tables 2 and 3  provides a PMC relation for each selected couple of bands and chemical composition. 

In particular  we find :
\small{
\begin{eqnarray}
\lefteqn{u=6.43-0.89\log{P}-3.78(u-g)-2.84\log{M}+0.36\log{Z}}\nonumber\\
\lefteqn{\sigma=0.14}\\ 
\lefteqn{g=-1.05-2.87\log{P}+3.35(g-r)-1.87\log{M}-0.06\log{Z}}\nonumber\\
\lefteqn{\sigma=0.02}\\   
\lefteqn{g=-0.83-2.87\log{P}+2.27(g-i)-1.94\log{M}-0.024\log{Z}}\nonumber\\
\lefteqn{\sigma=0.02}\\   
\lefteqn{g=-0.67-2.91\log{P}+1.99(g-z)-1.90\log{M}}\nonumber\\
\lefteqn{\sigma=0.07}
 \end{eqnarray}}
\noindent
for fundamental models and
\small{
\begin{eqnarray}
\lefteqn{u=5.77-1.48\log{P}-3.87(u-g)-2.72\log{M}+0.25\log{Z}}\nonumber\\
\lefteqn{\sigma=0.07} \\
\lefteqn{g=-1.53-3.11\log{P}+3.57(g-r)-1.63\log{M}-0.042\log{Z}}\nonumber\\ 
\lefteqn{\sigma=0.01}\\
\lefteqn{g=-1.279-3.07\log{P}+2.334(g-i)-1.72\log{M}-0.017\log{Z}} \nonumber\\
\lefteqn{\sigma=0.009}\\
\lefteqn{g=-1.151-3.099log{P}+2.02(g-z)-1.77\log{M}-0.0045\log{Z}}\nonumber\\
\lefteqn{\sigma=0.007}
\end{eqnarray}}
\noindent
for the first overtone ones.
As discussed in D04 for the Johnson-Cousins bands, these relations allow us to derive an estimate of the stellar mass for
RR Lyrae of known distance and color, while for cluster pulsators sharing the same distance and reddening, they provide direct estimates of the mass spread. 
On the other hand, if the mass and  colors are known, the same relations  can be used to infer individual and/or mean distance moduli of RR Lyrae stars in a given globular cluster or galaxy.

\subsection{The pulsation amplitudes}
\begin{figure*}
\centerline{
\includegraphics[width=12cm]{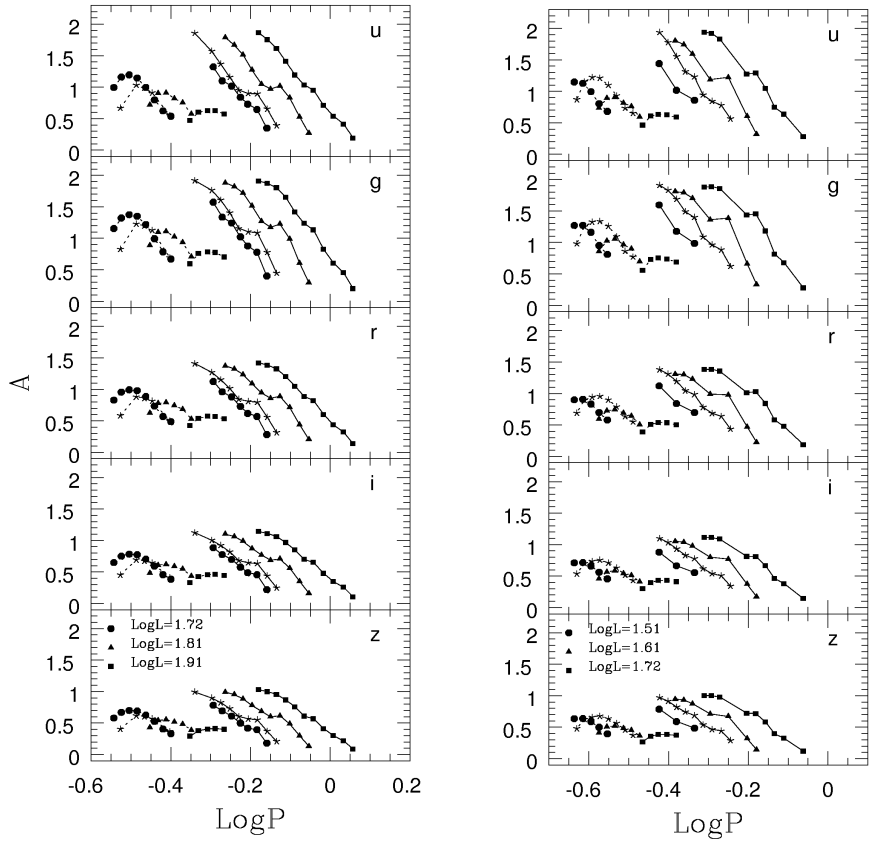}}
\caption{Period-amplitude diagram for FO (dashed line) and F (solid line)
  models with $M=080M_{\odot}$and Z=0.0001(left panel) and $M=0.65M_{\odot}$
  Z=.001(right panel) and for the three labelled values of $\log L/ \log
  L_{\odot}$. In both cases star symbols represent models with
  $M=0.75M_{\odot}$.}
\label{f9}
\end{figure*}

The pulsation amplitudes of the fundamental and first overtone multiwavelenght lightcurves are reported in  Tables 7 and 8. Their behaviour as a function of the pulsation 
period is shown in Fig. \ref{f9} for selected model sequences at $Z=0.0001$ (left panels) and $Z=0.001$ (right panels).We notice that the fundamental pulsators follow a linear behaviour, thus allowing the derivation of a mass dependent linear relation between the period, the
 pulsation amplitude and the absolute magnitude. The coefficients of these relations for the whole fundamental model set and the various photometric bands are reported in Table 6.
We notice that the coefficients of these relations are  expected to depend on the
adopted $\alpha$ parameter (see also D04). In particular in the \emph{g} filter the zero point and the amplitude coefficient  should vary by  
$\sim$ 0.14 and $\sim$ 0.38 respectively, as $\alpha$ increases from 1.5 to 2.0.

\begin{table}
 \centering
 %\begin{minipage}{15cm}
 \caption[]{Analytical coefficients of the PLA relations in the
 form $\log{P}=a+bA_i+c<M_i>+d \log{\frac{M}{M_{\odot}}}+e\log{Z}$.\label{tab4}}
 \begin{tabular}{ccccccc}
\hline
 $M_i$  &   $a$  &   $b$ &  $c$ & $d$ & $e$ & $\sigma$ \\
\hline
\hline
   u   &0.72  &-0.186  &-0.41  &-0.73 &0.038&0.02\\
   g   &0.14  &-0.184  &-0.371  &-0.54 &0.009&0.02\\
   r   &-0.03 &-0.174  &-0.372  &-0.60 &-0.015&0.01\\
   i   &-0.07 &-0.171  &-0.370  &-0.64 &0.069 &0.01\\
   z   &-0.074 &-0.164  &-0.368  &-0.66 &0.074&0.009\\
\hline
\end{tabular}
 %\end{minipage}
 \end{table}

\begin{figure*}
\centerline{
\includegraphics[width=12cm]{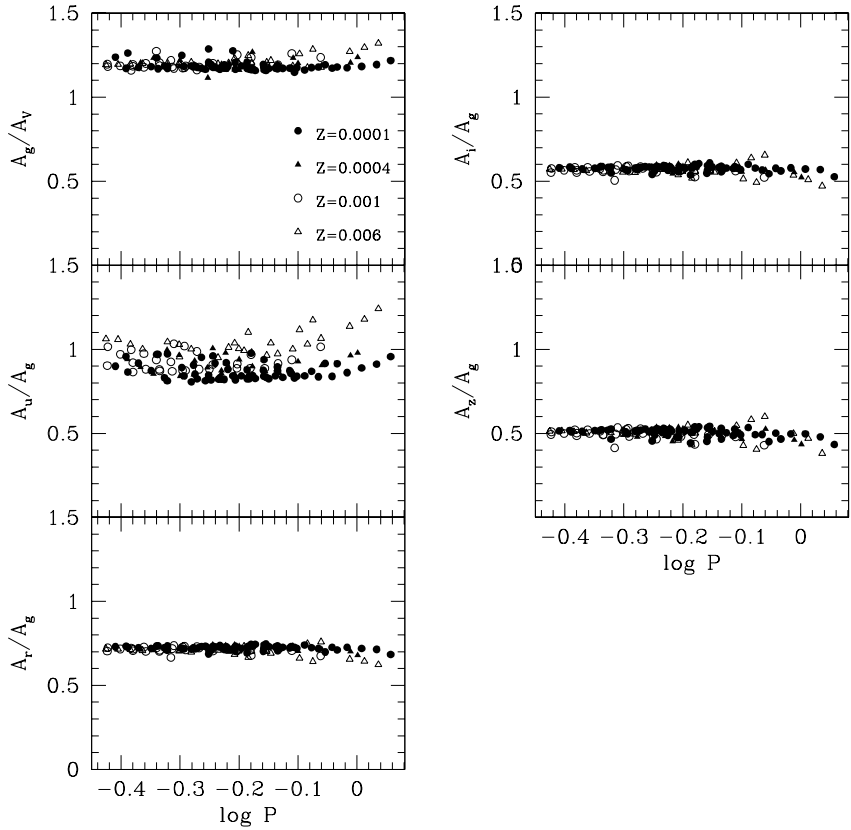}}
\caption{Amplitude ratios between the Johnson V and the SDSS \emph{g} bands
  (left top panel) and between the \emph{u, r, i, z} and the \emph{g} bands
  for fundamental  models.}
\label{f10}
\end{figure*}

\begin{figure*}
\centerline{
\includegraphics[width=12cm]{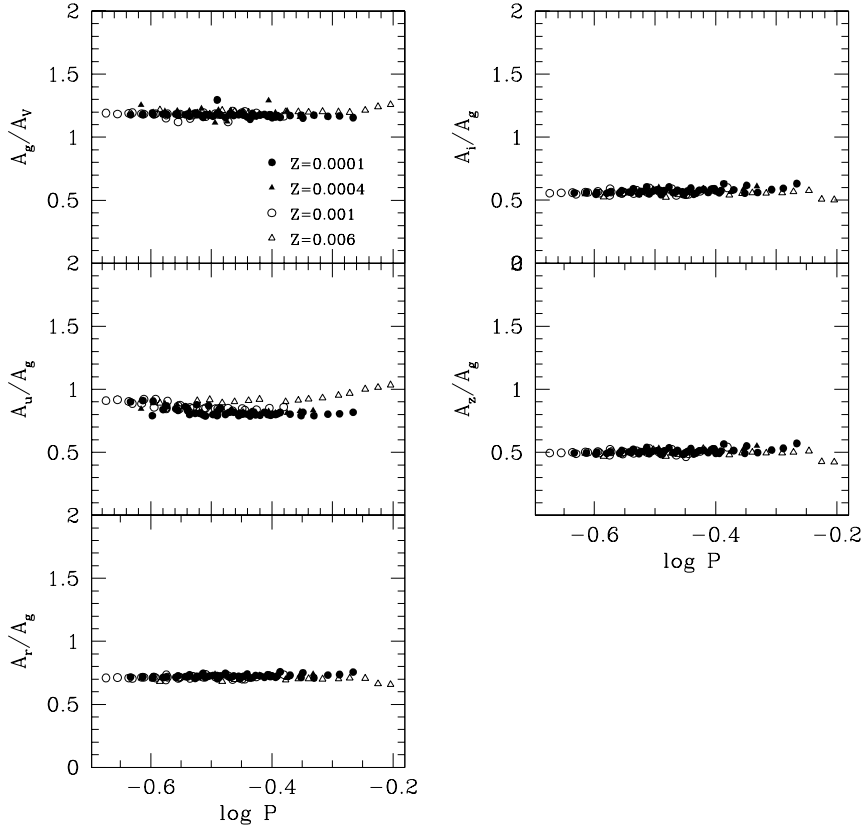}}
\caption{The same as the previous  figure but for first overtone models}
\label{f11}
\end{figure*}

As for first overtone pulsators we notice the characteristic bell shape also
shown in the bolometric and Johnson-Cousins band Bailey diagram (see e.g. \citealt{boncap}).  In Figs. \ref{f10} and \ref{f11} we show the amplitude ratios between the
Johnson V and the SDSS \emph{g} bands (left top panel of each figure) and
between the \emph{u, r, i, z} and the \emph{g} bands, for fundamental and first overtone
models respectively. We notice that the \emph{g} band amplitude is systematically
higher than the V amplitude, independently of the period and the adopted metal
abundance (see labels). At the same time the amplitudes in the \emph{r,i, z} filters
scale with a constant mean ratio (ranging from about 0.7 to 0.5 from \emph{r} to \emph{z})
with the \emph{g} band amplitude, thus suggesting that only few points along the
lightcurves in these filters will be required, if the g curve is accurately
sampled.  As for the \emph{u} band, the scatter is larger due to the significant
dependence on the adopted metallicity. This occurrence is more evident for fundamental models,
which are characterized by more asymmetric light curves and higher pulsation
amplitudes.

\begin{table*}
 \centering
 %\begin{minipage}{15cm}
 \caption[]{Pulsation amplitudes for the full  set of fundamental models (the full table is available in the electronic form).\label{tab1}}
 \begin{tabular}{cccccccccc}
\hline
 $Z$   &   $M/M_{\odot}$ &  $\log L/ \log L_{\odot}$ & $T_e$ (K) & P (d) & $A_u$ & $A_g$& $A_r$& $A_i$ & $A_z$ \\
\hline
\hline
0.0001  &   0.65  &   1.61   &   6900 &    0.4059  & 1.7877 &  1.8675 &  1.3689  & 1.0900  & 0.9669\\  
0.0001  &   0.65  &   1.61   &   6800 &   0.4260  & 1.6424 &  1.7906 &  1.2955  & 1.0261  & 0.9187\\
  ...   &  ...  &   ... &  ... &... &  ... &  ....  & ... &  ... &  ... \\
\hline
 \end{tabular}                                    
 %\end{minipage}                                  
 \end{table*}   
      
\begin{table*}
 \centering
 %\begin{minipage}{15cm}
 \caption[]{Pulsation amplitudes for the full  set of first overtone  models (the full table is available in the electronic form).\label{tab1}}
 \begin{tabular}{cccccccccc}
\hline
 $Z$   &   $M/M_{\odot}$ &  $\log L/ \log L_{\odot}$ & $T_e$ (K) & P (d) & $A_u$ & $A_g$& $A_r$& $A_i$ & $A_z$ \\
\hline
\hline
0.0001 &   0.65 &   1.61 &    7300& 0.2527 &  0.3939&  0.4978&  0.3522&  0.2727&  0.2427 \\
0.0001 &   0.65 &   1.61 &    7200& 0.2634 &  0.9037&  1.0765&  0.7677&  0.5969&  0.5296 \\
...   &  ...  &   ... &  ... &... &  ... &  ....  & ... &  ... &  ... \\
\hline
 \end{tabular}                                                                                           
 %\end{minipage}                                                                                         
 \end{table*}                                                                                            

We note that, once known the metal content, the mass term contained in  all the analytical relations reported in this Section can be replaced  with the value predicted by evolutionary horizontal branch computations at the selected metallicity
(see e.g. columns 5 and 6 of Table 1 in \citealt{boncap03}).

\section{Comparison with the observations}
In this section the predicted color-color loops in the various filter
combinations are compared with observed RR Lyrae samples. In particular we
test our predictions with the QUEST survey (see e.g. \citealt{viva01}) and the observations in the Draco galaxy (\citealt{bona}). The first one, a large census of field RR Lyrae, should map the history of
the outer halo of our Galaxy, the second should represent an example of RR Lyrae in a different enviroment.

\begin{figure}
\includegraphics[width=8cm]{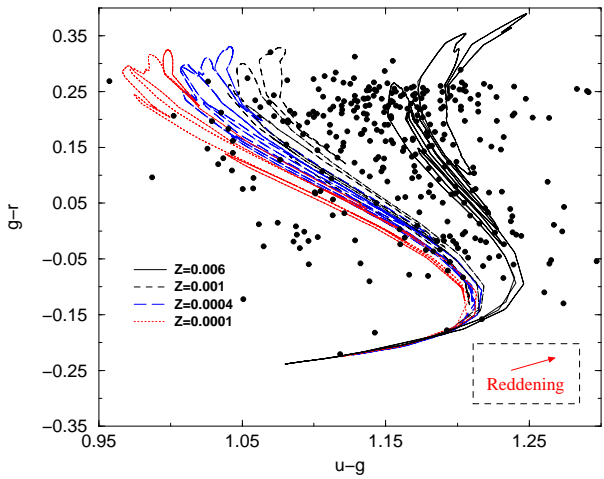}
\caption{Comparison between the theoretical loops shown in Fig. \ref{f6} and
  the QUEST RR Lyrae data (see text for details).}
\label{f12}
\end{figure}

\begin{figure}
\includegraphics[width=8cm]{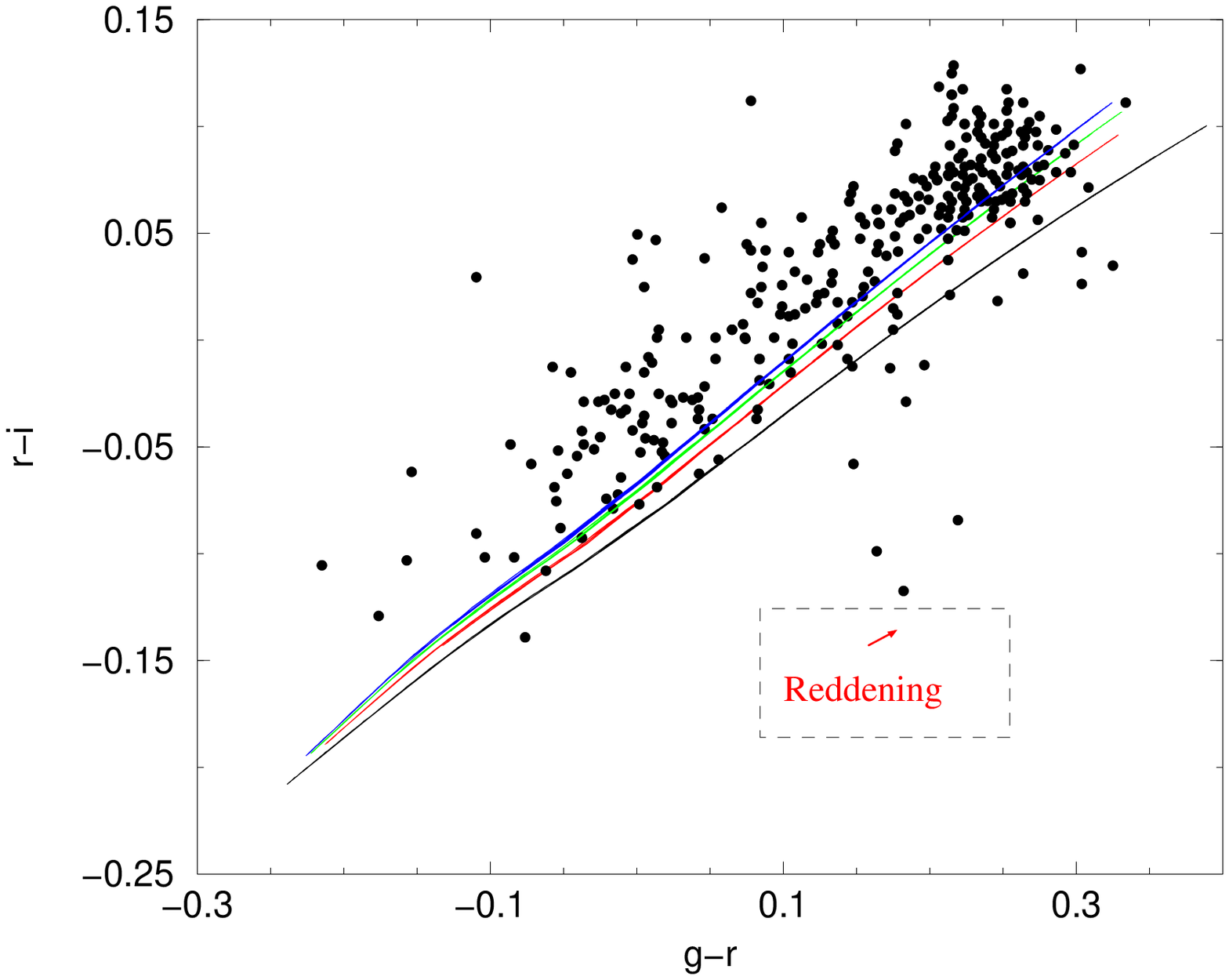}
\caption{Comparison between the theoretical loops shown in Fig. \ref{f7} and the
  QUEST RR Lyrae data (see text for details).}
\label{f13}
\end{figure}

\begin{figure}
\includegraphics[width=8cm]{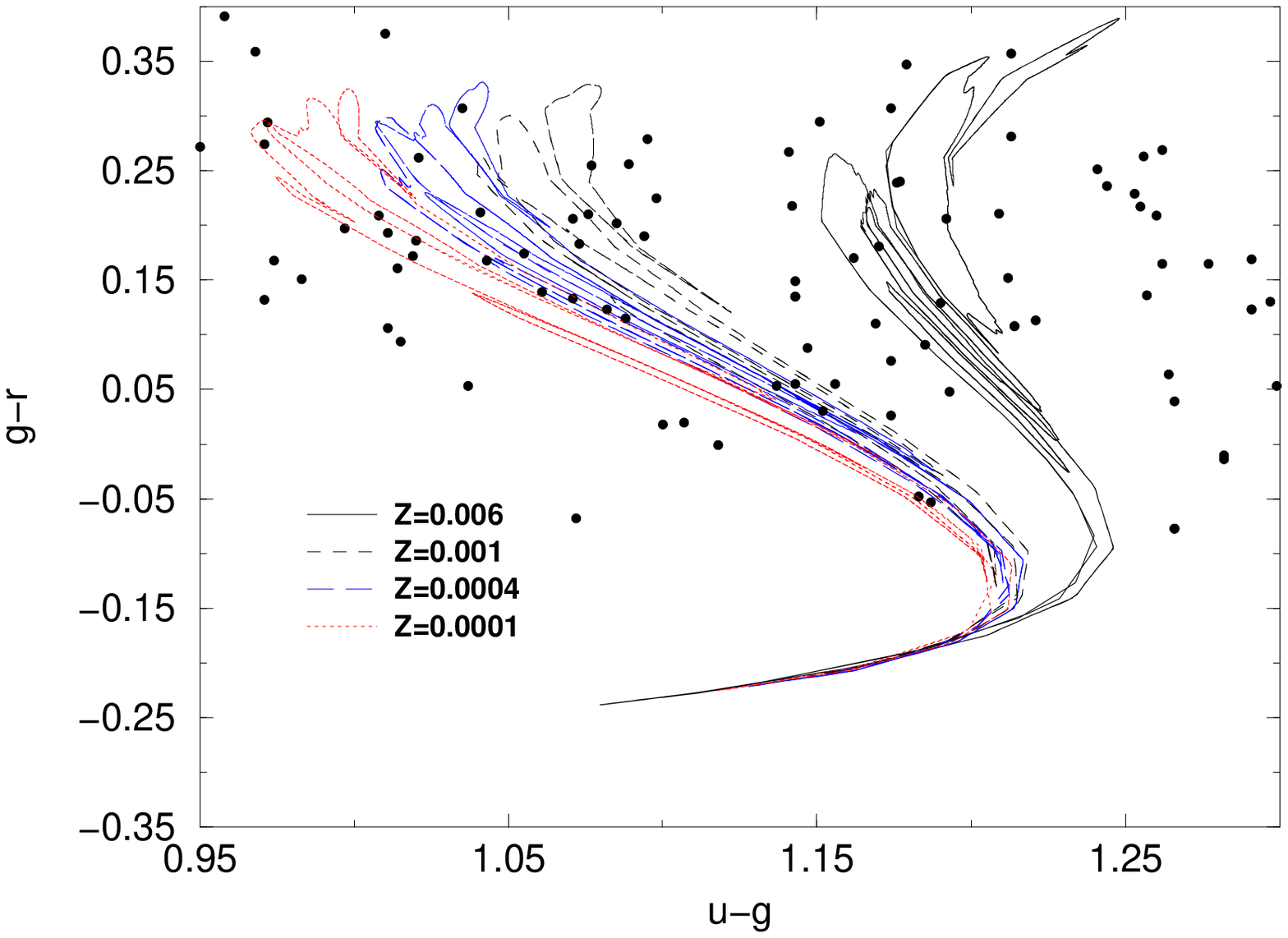}
\caption{The same of Fig. \ref{f12} but for Draco RR Lyrae.}
\label{f14}
\end{figure}

\begin{figure}
\includegraphics[width=8cm]{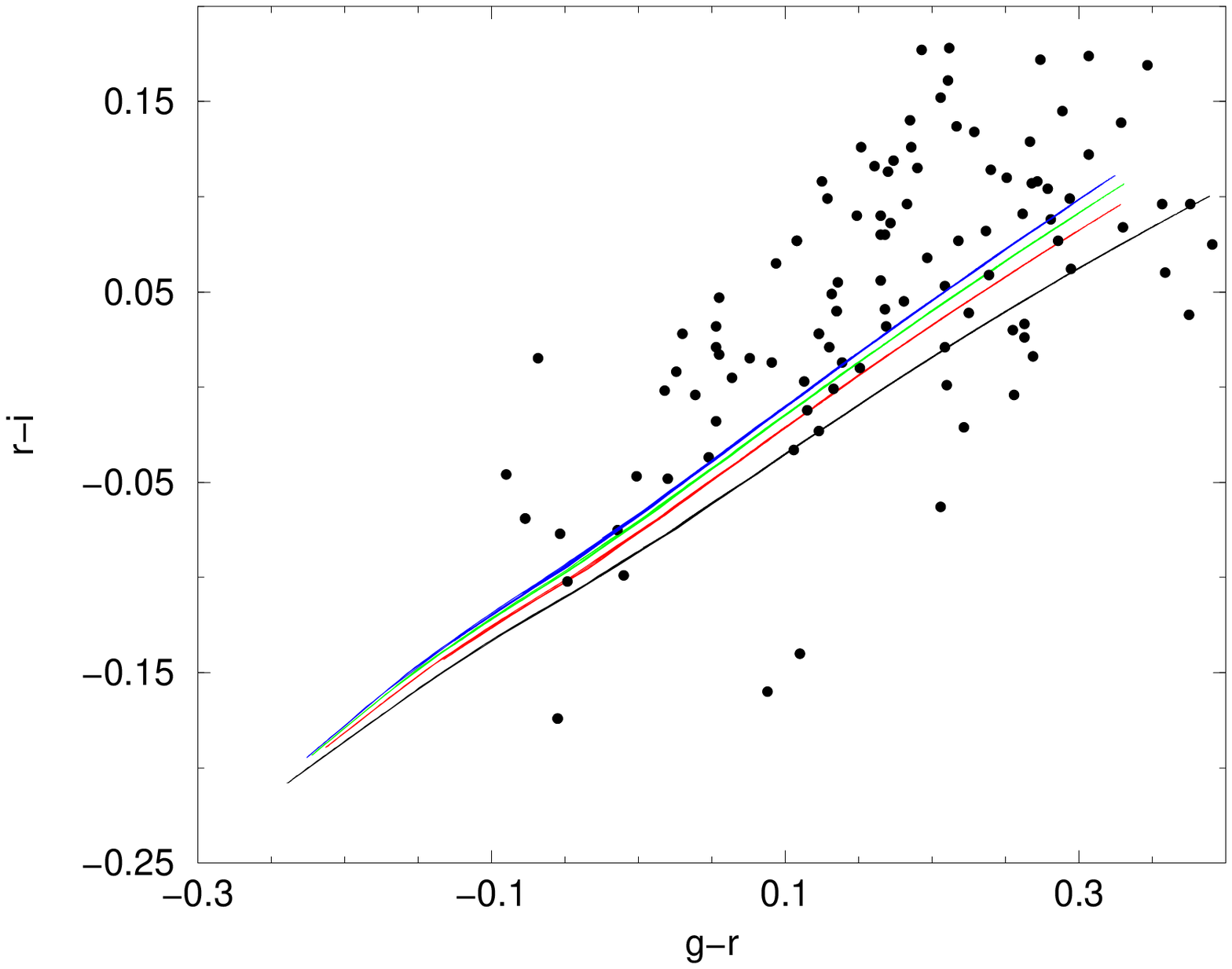}
\caption{The same of Fig. \ref{f13} but for Draco RR Lyrae.}
\label{f14a}
\end{figure}

Both the samples are positionally matched against the SDSS-DR4 (\citealt{adel}) catalogue using a search radius of 0.1 arcsec and subsequently dereddened (using IR maps, see \citealt{sch}).
Figs. \ref{f12} and \ref{f13} show the distributions  for QUEST data,
as observed in the \emph{g-r} vs \emph{u-g} and \emph{r-i} vs \emph{g-r} plane respectively. Superimposed, we show the theoretical loops for the labeled metallicities instead of average values because SDSS observations consist of  one or, at most, two phase points. 
The QUEST data which are expected to trace metal poor ($Z<0.001$) and distant halo stars (see \citealt{viva01}) are not matched by the corresponding model
loops, with the discrepancy larger than the mean uncertainty resulting from photometric ($\sigma (u-g) < 0.03$ mag) and reddening errors (less than 0.01 mag in colors, see also \citealt{ive05}). 
In Fig. \ref{f14} we show the comparison with DRACO RR Lyrae in the  \emph{g-r} vs \emph{u-g} plane. In this case we have a large spread in \emph{u-g} due in part to the significant photometric uncertainties ($\sigma (u-g) \sim 0.1$ mag and $\sigma (g-r) \sim 0.02$) affecting the data. In this case, the comparison with the theoretical loops does not allow us to discriminate a metallicity effect. However, the mean metallicity of this galaxy is generally considered poorer than Z=0.001 (see e.g. \citealt{mat}), thus it is noteworty that a consistent fraction of the Draco RR Lyare is redder than our Z=0.006 models with an overdensity at \emph{u-g} $> 1.25$ (difficult to explain with the photometric error alone). 
{\bf The situation is similar in the \emph{r-i} vs \emph{g-r} plane (see Fig. \ref{f14a})}.
%Even if an important factor determining the large dispersion in u-g color is the metallicity, the best agreement obtained for the Z=0.006 models . 
Possible sources for the discrepancies between theory and observation may result from observational uncertainties, including reddening and contamination effects, or theoretical biases, as for example the adopted model atmospheres and chemical mixture.

As for the reddening, a critical point is represented by the extinction law
adopted to relate the measured E(B-V) to the extinctions in the SDSS filters
(see e.g. \citealt{gir2004}). Assuming the coefficents tabulated by \citet{gir2004}, an underestimation of the E(B-V) color excesses of the order of 0.02 mag would require a redshift in the theoretical u-g and g-r of about 0.03 and 0.02 respectively, in the direction of reducing the discrepancy between data and metal-poor predictions in Figs. \ref{f12} and \ref{f13} (see the arrow in these plots).

\begin{figure}
\includegraphics[width=8cm]{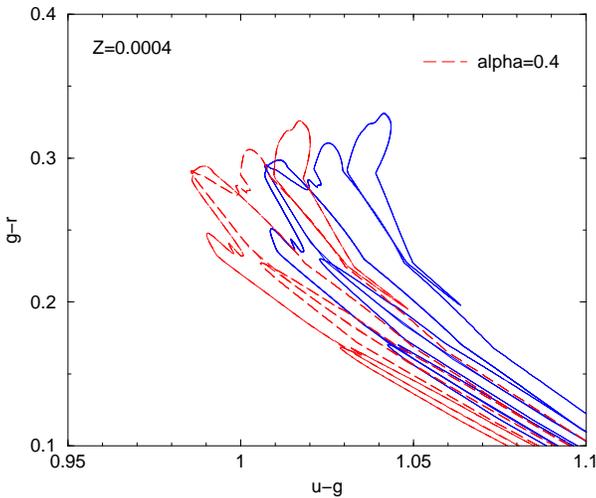}
\caption{Variation of the theoretical loops in the g-r vs u-g plane for models
  with Z=0.0004, as the $[\alpha/Fe]$ increases from 0 to 0.4.}
\label{f15}
\end{figure}

\begin{figure}
\includegraphics[width=8cm]{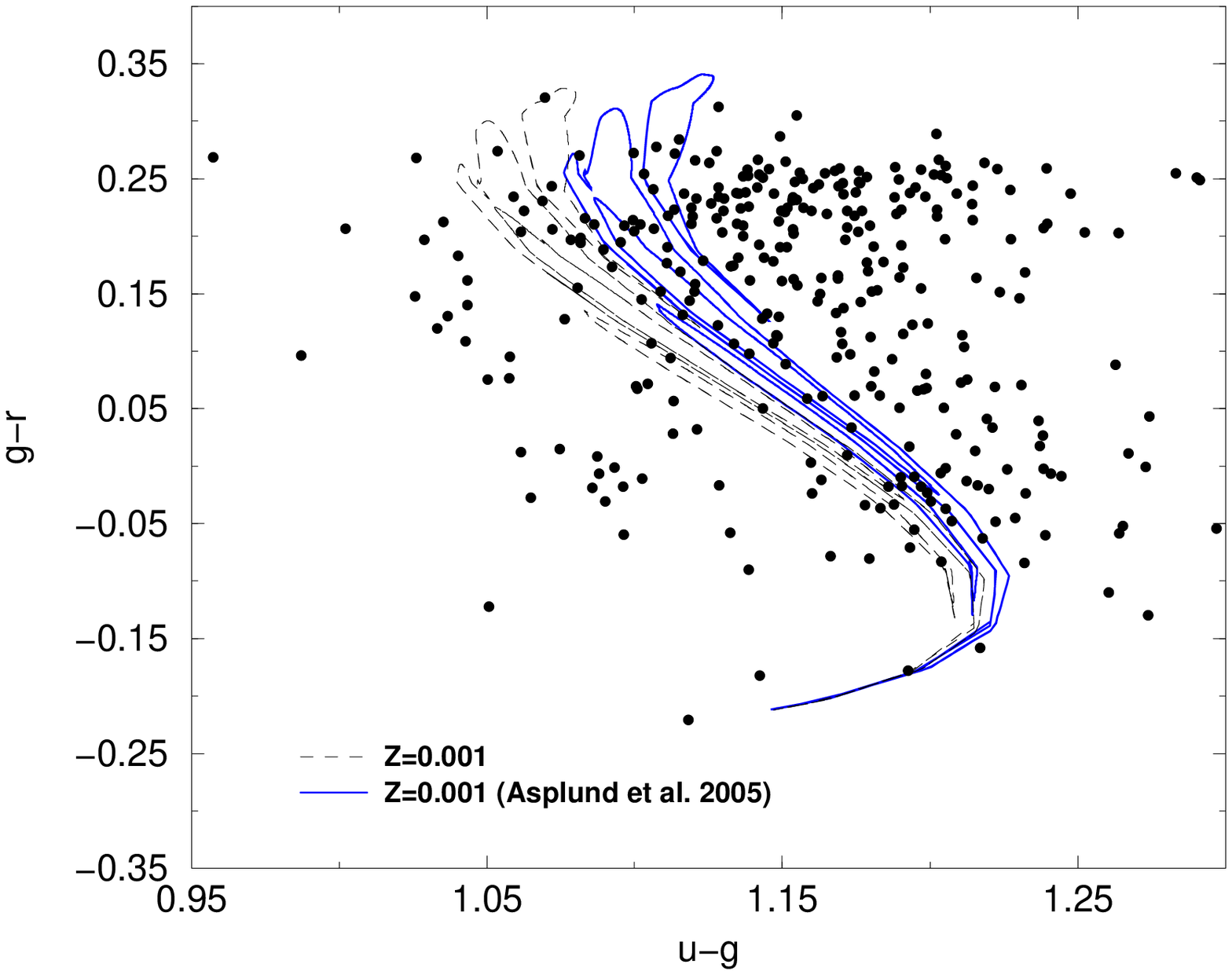}
\caption{Comparison between the theoretical loops shown in Fig. \ref{f6} and the
  QUEST RR Lyrae data (see text for details).}
\label{f16}
\end{figure}

Concerning model atmospheres, as discussed in section 3, we have adopted ATLAS
models \citep{castelli03} without $\alpha$ enhancement. In order to
explore the effect of possible $\alpha$ enhancement for the lowest
metallicities, as empirically suggested by various authors in the literature
(see e.g. \citealt{grat03}), we have transformed again the
theoretical bolometric light curves by using ATLAS model atmospheres with
$[\alpha/Fe]=0.4$ and reducing the adopted $[Fe/H]$ in order to obtain the
same global model metallicity (see \citealt{sala}). As an example, in
Fig. \ref{f15} we show the variation of the theoretical loops for models with Z=0.0004. It
appears that increasing $[\alpha/Fe]$ from 0 to 0.4 produces a larger
discrepancy between metal-poor models and the data. On the other hand, if in
the conversion from Z to $[Fe/H]$ we
replace the assumed solar metallicity with the more updated value by
\citet{asplu}, loops of a given metallicity move redward by about 0.05 mag in
\emph{g-r} vs \emph{u-g} plane (see e.g. figure \ref{f16} for Z=0.001), so that the recovered global metallicity for halo RR
Lyare is closer to typical values in the literature. 

\begin{figure}
\includegraphics[width=8cm]{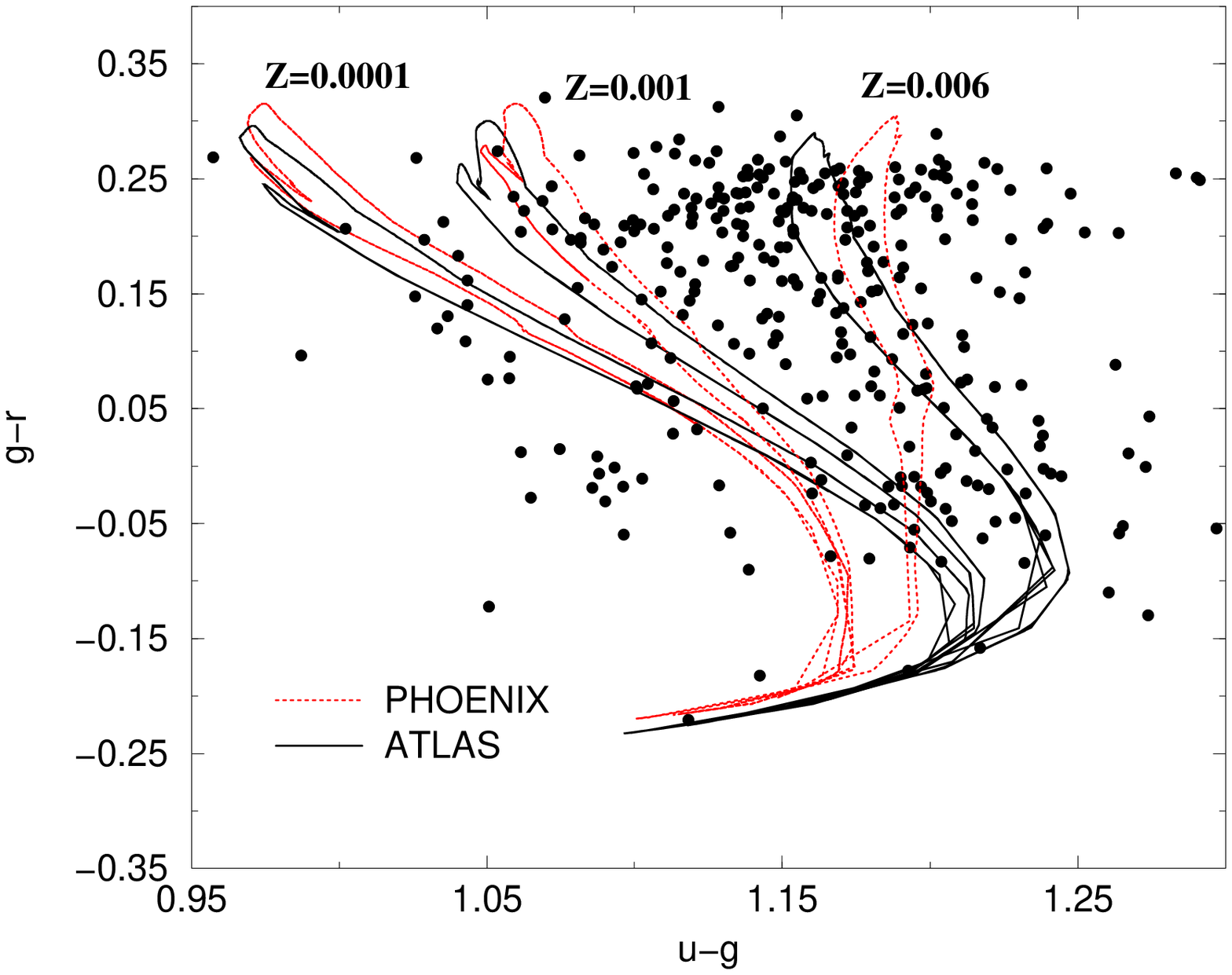}
\caption{The QUEST RR Lyrae  data compared with selectd models transformed
  both with ATLAS (solid lines) and PHOENIX (dotted lines) model atmospheres.}
\label{f17}
\end{figure}

To take into account the effect of the adopted set of model atmospheres, Fig. \ref{f17} plots the same comparison as in Fig. \ref{f12}  but using the PHOENIX atmospheres for giant stars (\citealt{kuci2006}), which have the main advantage of assuming the spherical geometry instead of the classical plane-parallel structure. From the u-g vs g-r plot, it is evident that the PHOENIX atmospheres affect the u-g color with a blueward shift (of about 0.03 mag) for models with g-r between 0.1 and 0.3 mag, corresponding to the bulk of the observed RR Lyrae. Moreover, the PHOENIX atmospheres lead to redder u-g colors (by about 0.05 mag) for stars with g-r lower than 0.1 mag, where the RR Lyare density is lower. However, both these effects produce minor changes in the behaviour observed in Figs. \ref{f12} and \ref{f13}.

\section{Conclusions}

We have transformed the predicted scenario for RR Lyrae stars with Z in the
range 0.0001-0.006 into the SDSS photometric system, providing theoretical
tools for the interpretation of modern large RR Lyrae surveys in these bands.
Mean magnitudes and colors and the pulsation amplitudes are used to derive
multiband analytical relations that can be used to constrain both the distances and the
stellar masses. Rather constant amplitude ratios are found for \emph{r,i, z} with
respect to the \emph{g} band, suggesting that only few points along the lightcurves 
in these filters will be required, if the g curve is accurately sampled.
The theoretical \emph{g-r} vs \emph{u-g} and \emph{r-i} vs \emph{g-r} are compared with available data in
the literature. In particular, for field RR Lyrae stars from the QUEST survey the comparison seems
to suggest a higher mean metallicity than usually assumed. In order to
understand this occurrence several sources of systematic errors have been discussed.

\section*{Acknowledgments}
We thank P. G. Prada Moroni and S. Degl'Innocenti for useful comments and suggestions on  filter transformations. 
Financial support 
for this study was provided by MIUR, under the scientific project
``On the evolution of stellar systems: 
fundamental step toward the scientific exploitation of VST'' 
(P.I. Massimo Capaccioli). This project made use of computational 
resources granted by the ``Consorzio di Ricerca del Gran Sasso'' 
according to the ``Progetto 6: Calcolo Evoluto e sue applicazioni 
(RSV6) - Cluster C11/B''.


\begin{thebibliography}{99}

\bibitem[\protect\citeauthoryear{Adelman-McCarthy et al.}{2006}]{adel}
Adelman-McCarthy J.K. et al., 2006, ApJS, 162, 38

\bibitem[\protect\citeauthoryear{Alcal\'a et al.}{2006}]{alca}Alcal\'a J.M. et
al., 2006, MmSAI, in press.

\bibitem[\protect\citeauthoryear{Asplund, Grevesse \& Sauval}{2005}]{asplu}
    Asplund M., Grevesse N., \& Sauval, A.J. 2005, in ``Cosmic Abundances as
    Records of Stellar Evolution and Nucleosynthesis", eds. F.N. Bash, \&
    T.J. Barnes, ASP Conf. Series, 336, 25


\bibitem[\protect\citeauthoryear{Bonanos et al.}{2004}]{bona}Bonanos A.Z.,
Stanek K.Z., Szentgyorgyi A.H., Sasselov D.D., Bakos G.A., 2004, AJ, 127, 861

\bibitem[\protect\citeauthoryear{Bono, Castellani \& Marconi}{2000}]{bcm00}
  Bono G., Castellani V., Marconi M., 2000, ApJ Letters, 532, 129


\bibitem[\protect\citeauthoryear{Bono \& Stellingwerf}{1994}]{bonostell} Bono
G., Stellingwerf R. F., 1994, ApJS, 93, 233

\bibitem[\protect\citeauthoryear{Bono et al.}{1997}]{boncap} Bono G., Caputo
F., Castellani V., Marconi M., 1997, A\&AS, 121, 327


\bibitem[\protect\citeauthoryear{Bono et al.}{2003}]{boncap03}Bono G., Caputo
F., Castellani V., Marconi M., Storm J., Degl'Innocenti S., 2003, MNRAS, 344,
1097

\bibitem[\protect\citeauthoryear{Brocato, Castellani \& Ripepi}{1996}]{bcr96}
  Brocato E., Castellani V., Ripepi V., 1996, AJ, 111, 809

\bibitem[\protect\citeauthoryear{Brown et al.}{2004}]{bro}Brown R.H.  et al.,
2004, ApJ, 613, 125

\bibitem[\protect\citeauthoryear{Cacciari et al.}{2003}]{cacc}Cacciari C.,
2003, in G. Piotto, G. Meylan, S. G. Djorgovski, M. Riello, eds.  "New
Horizons in Globular Cluster Astronomy", ASP Conference Proceedings, Vol. 296,
329

\bibitem[\protect\citeauthoryear{Caputo et al.}{2000}]{capcast}Caputo F.,
Castellani V., Marconi M., Ripepi V., 2000, MNRAS, 316, 819


\bibitem[\protect\citeauthoryear{Castelli, Gratton \& Kurucz}{1997}]{castgrat}
  Castelli F., Gratton R. G., Kurucz R. L., 1997, A\&A, 318, 841

\bibitem[\protect\citeauthoryear{Castelli \& Kurucz}{2003}]{castelli03}
  Castelli F., Kurucz R. L., 2003, IAU Symp. 210, Modeling of Stellar Atmospheres,
ed. N. E. Piskunov, W. W. Weiss, \& D. F. Gray (San Francisco: ASP)

\bibitem[\protect\citeauthoryear{Di Criscienzo et al.}{2006}]{dicri06}Di Criscienzo M., Caputo F., Marconi M., Musella I., 2006, MNRAS, 365, 1357


\bibitem[\protect\citeauthoryear{Di Criscienzo, Marconi \& Caputo}{2004}]{dicri04}Di Criscienzo M., Marconi M., Caputo F., 2004, ApJ, 612, 1092

\bibitem[\protect\citeauthoryear{Dinescu et al.}{2004}]{dine}Dinescu D.I.,
Keeney B.A., Majewski S.R., Girard T.M., 2004, AJ, 128, 687

\bibitem[\protect\citeauthoryear{Fukugita}{1996}]{fuku}Fukugita M. et al., 1996, AJ, 111, 1748


\bibitem[\protect\citeauthoryear{Girardi et al.}{2002}]{gir2002} Girardi et
al., 2002, A\&A, 391, 195
\bibitem[\protect\citeauthoryear{Girardi et al.}{2004}]{gir2004}Girardi L.,
Grebel E. K., Odenkirchen M., Chiosi C., 2004, A\&A, 422, 205

\bibitem[\protect\citeauthoryear{Grevesse \& Sauval}{1998}]{greve98}Grevesse N., Sauval A. J., 1998, Space Science Reviews, 85, 161 


\bibitem[\protect\citeauthoryear{Gratton et al.}{2003}]{grat03}Gratton R.G.,
  Carretta E., Desidera S., Lucatello S., Mazzei P., Barbieri M., 2003, A\&A,
  406, 131




\bibitem[\protect\citeauthoryear{Ivezic et al.}{2005}]{ive05}Ivezic Z., Vivas A.K., Lupton R.H., Zinn R., 2005, AJ, 129, 1096


\bibitem[\protect\citeauthoryear{Ivezic et al.}{2000}]{ive00}Ivezic Z., et al., 2000, AJ, 120, 963


\bibitem[\protect\citeauthoryear{Kucinskas et al.}{2005}]{kuci2005}Kucinskas A. et al. 2005, A\&A, 442, 281


\bibitem[\protect\citeauthoryear{Kucinskas et al.}{2006}]{kuci2006}Kucinskas A. et al., 2006, astro-ph/0603416

 \bibitem[\protect\citeauthoryear{Kurucz}{1990}]{kur90}Kurucz R. L., 1990, Stellar Atmospheres: Beyond Classical Models, NATO Asi Ser., ed. L. Crivellari et al., 441


\bibitem[\protect\citeauthoryear{Lynden-Bell}{1982}]{lyn} Lynden-Bell D., 1982, Observatory, 102, 202




\bibitem[\protect\citeauthoryear{Marconi et al.}{2003}]{marc03}Marconi M., Caputo F., Di Criscienzo M., Castellani M., 2003, ApJ, 596, 299

\bibitem[\protect\citeauthoryear{Marconi et al.}{2006}]{marc06}Marconi M., et al.. 2006, MmSAI, in press

\bibitem[\protect\citeauthoryear{Mateo}{1998}]{mat} Mateo M. L., 1998, ARA\&A, 36, 435


\bibitem[\protect\citeauthoryear{Salaris, Chieffi \& Straniero}{1993}]{sala}Salaris M., Chieffi A., Straniero O., 1993, ApJ, 414, 580

 \bibitem[\protect\citeauthoryear{Schlegel, Finkbeiner \& Davis}{1998}]{sch}Schlegel D. J., Finkbeiner D. P., Davis M. 1998, ApJ, 500, 525 



 \bibitem[\protect\citeauthoryear{Van Albada \& Baker}{1971}]{vb71}Van Albada T.S., Baker N., 1971, ApJ, 169, 311

 \bibitem[\protect\citeauthoryear{Vivas et al.}{2001}]{viva01}Vivas A.K.,  et al., 2001, ApJ, 554, 33



\bibitem[\protect\citeauthoryear{Vivas et al.}{2004}]{viva04}Vivas A.K.,  et al., 2004, AJ, 127, 1158

\bibitem[\protect\citeauthoryear{Vivas et al.}{2006}]{viva06}Vivas A.K.,  Zinn R., 2006, astro-ph/0604359


\bibitem[\protect\citeauthoryear{Wu et al.}{2005}]{wu}Wu C., Qiu Y.L., Deng J.S., Hu J.Y., Zhao Y.H., 2005, AJ, 130, 1640


\end{thebibliography}
\end{document}